# Phase transitions induced by confinement of ferroic nanoparticles.

Anna N.Morozovska [*], Maya D.Glinchuk, Eugene A. Eliseev,

Institute for Problems of Materials Science, NAS of Ukraine,

Krjijanovskogo 3, 03142 Kiev, Ukraine, glin@materials.kiev.ua, eliseev@i.com.ua

General approach for consideration of primary ferroic (ferroelectric, ferromagnetic, ferroelastic) nanoparticles phase transitions was proposed in phenomenological theory framework. The surface stress, order parameter gradient, striction as well as depolarization and demagnetization effects were included into the free energy. The strong intrinsic surface stress under the curved nanoparticle surface was shown to play the important role in the shift of transition temperature (if any) up to the appearance of new ordered phase absent in the bulk ferroic.

The way of transition temperature, critical sizes and other properties detailed calculations was demonstrated on the example of conventional and incipient ferroelectric nanospheres and nanorods with perovskite structure. For this purpose Ginzburg-Landau-Devonshire free energy expansion on polarization and stress powers has been derived allowing for the mechanical conditions on nanoparticle surface. Obtained Euler-Lagrange equations were solved by direct variational method. This leads to the conventional form of the free energy with renormalized coefficients depending on nanoparticle sizes, surface stress and electrostriction tensor values, and so opens the way of polar properties calculations by algebraic transformations. Surface piezoeffect causes built-in electric field that induces electret-like polar state and smears the phase transition point.

The approximate analytical expression for the size-induced ferroelectric transition temperature dependence on cylindrical or spherical nanoparticle sizes, polarization gradient coefficient, correlation radius, intrinsic surface stress and electrostriction coefficient was derived.

It was shown that the transition temperature of nanoparticle could be higher than the one of the bulk material. The best conditions of ferroelectric properties conservation and enhancement in nanowires correspond to the radius 5-50nm and compressive surface stress. Under the favorable conditions size effects (spatial confinement) induces ferroelectric phase in incipient ferroelectrics nanowires and nanospheres. The prediction of size-induced ferroelectricity in $KTaO_3$ nanorods with radius less then 5-20 nm at room temperatures could be important for the next step of device miniaturization based on 3D nanostructures.

[*] Corresponding author, morozo@i.com.ua

permanent address: V. Lashkarev Institute of Semiconductor Physics, NAS of Ukraine,

41, pr. Nauki, 03028 Kiev, Ukraine

## 1. Introduction

Up to now phase transitions in solids attract much scientific and technical interest because of the properties anomalies in the vicinity of phase transition temperature. Recently the possibility to govern the appearance of phase transitions at any arbitrary temperature has been demonstrated in nanosized materials due to the so-called size-driven phase transition. Such transitions were observed in many solids, including ferroelectric, ferromagnetic and ferroelastic ones. Ferroelectric, ferromagnetic and ferroelastic materials are known to belong to the primary ferroics [1], because electric, magnetic or elastic field respectively could perform the switching of the order parameter. The common feature of nanomaterials with sizes less than 100nm, such as films and nanoparticles of different shape, is essential influence of the surface on their properties. Allowing for all surface properties (including symmetry, electronic, photonic etc) differs from those in the bulk, nanomaterials open the way to obtain a variety of new unique properties [2], a lot of which are useful for applications, e.g. the ability to store and release energy in well-regulated manners, making them very useful for sensors and actuators.

Among different ferroic nanomaterials magnetic ones are the most thoroughly investigated, in particular thin magnetic films and their multilayers [1, 3]. The outlook of ferroelectric thin films applications in memory devises leads to the intensive investigations of their polar and dielectric properties anomalies during last decade [4]. Recently the investigations of the cylindrical and spherical ferroic nanoparticles became a hot topic, because of the properties new behavior, absent in the bulk. For instance, room temperature ferromagnetism has been observed in spherical nanoparticles (size 7-30nm) of nonmagnetic oxides such as $CeO_2$, $Al_2O_3$, $ZnO$ *etc* [2]. Extremely strong superparamagnetic behavior down to 4K has been found in gold and palladium nanoparticles, which are nonmagnetic in the bulk [5]. Particles of both metals had a narrow size distribution with a mean diameter of 2.5nm. To the best of our knowledge the quantitative explanation and theoretical description of size-driven magnetism in nanoparticles are absent.

Keeping in mind the similarity of the ferroic properties, one could expect the appearance of ferroelectricity in highly-polarizable paraelectric nanoparticles induced by spatial confinement. Unfortunately, nothing of this kind has been revealed up to now. The only encouraging result has been recently reported by Yadlovker and Berger [6]. They reported about the polarization enhancement and ferroelectric phase conservation in Roshelle salt nanorods of diameter 30nm up to material decomposition temperature $55^0C$, that is about $30^0C$ higher than the transition temperature of the bulk crystals. The phenomenological description of ferroelectricity enhancement in confined nanorods has been recently proposed [7], [8].

To check the possibility of ferroelectricity appearance in nanoparticles of the materials, which are nonferroelectric in the bulk, as well as to reveal the common features responsible for the new phases appearance in primary ferroic nanoparticles, in the paper we study phase transitions in them. The calculations have been performed in the phenomenological theory framework for the case of single-domain ferroics, that is known to be valid for small enough sizes (less than 100 nm) [1], [9], [10], [11], [12], [13]. We took into account the contribution of the surface stress into the free energy, the gradient of order parameter (magnetization, polarization or strain), as well as depolarization, demagnetization or de-elastification effects, since all these factors are known to influence strongly on the phase transitions in nanomaterials. However, in the most of the theoretical papers, devoted to the consideration of the size effects in spatially confined systems the simultaneous influence of aforementioned factors and especially surface stress on the phase transitions was neglected (see e.g. [2], [14], [15], [16], [17], [18]).

Here we have shown that phase transitions absent in the bulk appear in cylindrical or spherical ferroic nanoparticles under the favorable conditions. The detailed consideration was performed for incipient ferroelectric $KTaO_3$, that is paraelectric up to zero Kelvin in the bulk. The theory predicts optimal sizes for ferroelectricity appearance in incipient ferroelectric nanoparticles.

## 2. Basic concepts for ferroic nanoparticles

When elaborating thermodynamic theory for the description of surface and size effects on polar properties and phase diagrams of ferroelectric nanoparticles of different shape, we will use Ginsburg-Landau-Devonshire phenomenological approach (see e.g. [19], [20], [21], [22]) with respect to the surface energy, correlation (gradient) energy, depolarization field and mechanical stress.

Characteristic feature of the nanoscale structures phenomenological description is the surface energy contribution that becomes comparable with the bulk one and can exceed it under size decrease.

For the case of primary ferroics Landau-Ginzburg-Devonshire expansion of bulk ($G_V$) and surface ($G_S$) parts of Hibbs free energy on the multicomponent order parameter $\boldsymbol{\eta}$ powers (vectors of polarization, magnetization or strain tensor for ferroelectric, ferromagnetic or ferroelastic media respectively) and stress tensor components powers $\hat{\sigma}$ has the form:

$$G_V = \int_V d^3r \left( \begin{array}{c} \dfrac{a_{ij}(T)}{2}\eta_i\eta_j + \dfrac{a_{ijkl}}{4}\eta_i\eta_j\eta_k\eta_l + \dfrac{a_{ijklmn}}{6}\eta_i\eta_j\eta_k\eta_l\eta_m\eta_n + ... + \\ + \dfrac{g_{ijkl}}{2}\left(\dfrac{\partial \eta_i}{\partial x_j}\dfrac{\partial \eta_k}{\partial x_l}\right) - \eta_i\left(E_{0i} + \dfrac{E_i^d}{2}\right) - Q_{ijkl}\sigma_{ij}\eta_k\eta_l - \dfrac{1}{2}s_{ijkl}\sigma_{ij}\sigma_{kl} \end{array} \right), \quad (1a)$$

$$G_S = \int_S d^2r \left( \begin{array}{c} \dfrac{a_{ij}^S}{2}\eta_i\eta_j + \dfrac{a_{ij}^S}{4}\eta_i^2\eta_j^2 + \dfrac{a_{ijk}^S}{6}\eta_i^2\eta_j^2\eta_k^2 - q_{ijkl}^S\sigma_{ij}\eta_k\eta_l \\ + d_{ijk}^S\sigma_{jk}\eta_i + \mu_{\alpha\beta}^S s_{\alpha\beta jk}\sigma_{jk} + \dfrac{v_{ijkl}^S}{2}\sigma_{ij}\sigma_{kl} + ... \end{array} \right). \quad (1b)$$

Coefficients $a_{ij}(T)$ explicitly depend on temperature $T$ in the framework of Landau-Ginzburg-Devonshire approach. Coefficients $a_{ij}^S$ of the surface energy expansion may also depend on temperature.

High order expansion coefficients $a_{ijkl}$, $a_{ijklmn}$, $a_{ijkl}^S$ and $a_{ijklmn}^S$, are supposed to be temperature independent, constants $g_{ijkl}$ determine magnitude of gradient energy. Tensors $g_{ijkl}$ and $a_{ijklmn}$ are positively defined. Situation with tensor $a_{ijkl}$ depends on the phase transition order, namely tensor and $a_{ijkl}$ is positively defined for the second order phase transition, while for the first order ones it is negatively defined. $\mathbf{E}_0$ is external field conjugated with order parameter $\mathbf{\eta}$.

$\mathbf{E}^d$ is depolarization, demagnetization or de-elastification field that increases due to the inhomogeneous distribution of order parameter $\mathbf{\eta}$ in confined system; it is easy to show that they are related to each other via linear operator $\hat{N}^d[\mathbf{\eta}]$ as $\mathbf{E}^d \cong \hat{N}^d[\mathbf{\eta}]$ [19]. In general case of spatial inhomogeneity of $\mathbf{\eta}$ operator $\hat{N}^d[\mathbf{\eta}]$ has only integral representation (see e.g. [23]). The field $\mathbf{E}^d$ tries to suppress ordered phase inside the system.

Coupling terms $Q_{ijkl}\sigma_{ij}\eta_k\eta_l$ and $q_{ijkl}^S\sigma_{ij}\eta_k\eta_l$ determine the influence of mechanical stress on the order parameter for the materials with high symmetry paraphase (paraelectric, paramagnetic or paraelastic). Here $Q_{ijkl}$ and $q_{ijkl}^S$ are respectively the bulk and surface striction coefficients; $s_{ijkl}$ are components of elastic compliance tensor [24]. The symmetry of surface striction tensor $q_{ijkl}^S$ is the same as bulk striction $Q_{jklm}$ one, but their signs and relative values can be different. For instance, anomalously large surface magnetostriction was observed in thin films NiFe/Ag/Si, NiFe/Cu/Si, and Ni/SiO$_2$ of thickness below about 5 nm [25].

In (1b) the surface piezoeffect tensor $d_{ijk}^S$ is introduced. It arises even in cubic paraelectrics due to the symmetry breaking near the surface (vanishing of inversion center, see e.g. [26], [27]), while in magnetics it exists, when there is no inversion of time among the symmetry operations of the material. Tensor $v_{jklm}^{Si}$ is related with the surface excess elastic moduli.

The intrinsic surface stress $\mu_{\alpha\beta}^S$ exists under the curved surface of solid body and determines the excess pressure on the surface [28], [29]. Surface stress tensor $\mu_{\alpha\beta}^S$ is defined as the derivative of the surface energy on deformation tensor. Let us underline, that in many experimental papers (e.g. [30], [31], [32]) size effects of ferroelectric nanoparticles phase diagrams are related with the intrinsic surface stress (or surface tension by analogy with liquids). Intrinsic mechanical stress under curved surface is determined by the tensor of intrinsic surface stress $\mu_{\alpha\beta}^S$:

$$n_k \sigma_{kj}\big|_S = -\frac{\mu_{\alpha\alpha}^S}{R_\alpha} n_j, \qquad (2)$$

where $R_\alpha$ are the main curvatures of surface free of facets and edges in continuum media approximation, $n_k$ are components of the external normal [28], [29]. In the case of mechanical stress homogeneous distribution $\hat{\sigma} = -\frac{\mu_{\alpha\alpha}^S}{R_\alpha} \hat{L}$, where $\hat{L}$ is the second rank tensor with constant coefficients. The form of tensor $\hat{L}$ is determined by the nanoparticle shape (e.g. spherical, ellipsoidal or cylindrical) and mechanical boundary conditions.

The sign of surface stress tensor components $\mu_{\alpha\beta}^S$ depends on the chemical properties of the nanoparticle ambient material, the presence of oxide or interface layer [29]. For the case of chemically pure surface, in thermodynamic equilibrium with inert environment, diagonal elements $\mu_{\alpha\alpha}^S$ should be positive similarly to the case of surface tension for liquids. Taking into account, that there exists surface layers/interfaces with chemical, structural and polar properties different from those of the bulk, hereinafter we consider both positive and negative values of $\mu_{\alpha\alpha}^S$.

For the considered case of nanoparticles with diameter less than 100nm, stress $\hat{\sigma}$ can be considered as homogeneous. Its contributions into the renormalization of the quadratic terms coefficients in Eqs.(1) via striction effect is the following:

$$a_{Rij}^S = \left(a_{ij}^S(T) + 2q_{lkij}^S L_{lk} \frac{\mu_{\alpha\alpha}^S}{R_\alpha}\right), \qquad a_{Rij} = \left(a_{ij}(T) + 2Q_{lkij} L_{lk} \frac{\mu_{\alpha\alpha}^S}{R_\alpha}\right). \qquad (3)$$

For the conventional ferroics $a_{ij}(T)$ have the view $a_{ij}(T) = \delta_{ij}\alpha_T(T - T_C^i)$, where $T_C^i$ is the Curie temperature of the bulk material for the order parameter component $\eta_i$. Neglecting of $\mathbf{E}^d$ and gradient contribution, one obtains from Eq.(3) the Curie temperature renormalization:

$$T_{CR}^x = T_C^x - \frac{2\tau_x}{\alpha_T}\frac{\mu_{\alpha\alpha}^S}{R_\alpha}, \quad T_{CR}^y = T_C^y - \frac{2\tau_y}{\alpha_T}\frac{\mu_{\alpha\alpha}^S}{R_\alpha}, \quad T_{CR}^z = T_C^z - \frac{2\tau_z}{\alpha_T}\frac{\mu_{\alpha\alpha}^S}{R_\alpha}, \qquad (4)$$

where the constants $\tau_i = Q_{lkii}L_{lk}$ ($i = x, y, z$) are introduced. It is seen that renormalized Curie temperatures $T_{CR}^i$ are different for order parameter components $\eta_i$, since the constants $\tau_i$ can be different depending on subscript $i$. The shifts of $T_{CR}^i$ originated from surface stress $\hat{\sigma}$ lead not only to the change of the nanoparticle phase diagram in comparison with a bulk ferroic system, but under the favorable conditions (e.g. at $\tau_i \mu_{\alpha\alpha}^S < 0$ and $T_{CR}^i > 0$) to the appearance of the new phases absent in the bulk. In particular case when cubic bulk system transforms into the polar tetragonal phase with six possible orientations of order parameter (like ferroelectric $PbTiO_3$ or multiferroic $BiFeO_3$) at $T < T_C^b$, the confined system subjected to the surface stress of arbitrary symmetry could transform into the polar phase with only two possible orientations of order parameter (e.g. $\pm\eta_3$) corresponding to the component with the highest transition temperature ($T_{CR}^z$).

However in order to obtain rigorous renormalization of Curie temperature one should take into consideration the renormalization of $a_{ij}(T)$ originated from the inner field $\mathbf{E}^d$ and order parameter gradient, so Eqs.(3-4) are valid only for the case when the surface stress contribution is much larger than the terms originated from $\mathbf{E}^d\boldsymbol{\eta}$ and $g_{iiii}(\nabla\eta_i)^2$. It is not excluded, since the surface stress $\sigma$, existing in nanoparticles with radius of curvature $R = 5-50$ nm, is about $10^8 - 10^{10}$ Pa at the reasonable values of surface stress tensor $|\mu_{\alpha\alpha}^S| = 5-50$ N/m [29], [31]. Therefore the stress induced by surface curvature is very strong and so it may cause noticeable shift of the bulk phase transition temperature (if any).

Field $\mathbf{E}^d$ always leads to the decrease of Curie temperature. However, it vanishes in some important cases, e.g. long nanorods with order parameter aligned along the cylinder axis [7], [8] or magnetic particles with superconducting covering.

For the sake of rigorous consideration of all contributions into Curie temperature, let us perform calculations for ferroic materials with definite characteristics. Namely, we will consider size-induced transitions between paraphase and ordered phases for the one component order parameter in cylindrical and spherical perovskite nanoparticles of conventional and incipient ferroelectrics.

To be sure that our efforts will not be in vain, let us perform simple estimations. It is obvious from Eq.(4) that the transition temperature shift depends on $Q_{lkii}L_{lk}$ values and signs, and it increases with the particle radius decrease. Typical bulk electrostriction coefficients $Q_{lkij}$ are order of magnitude $0.1-0.05$ m$^4$/C$^2$, that leads to the estimation of surface stress via striction contribution into $a_{ij}(T)$ renormalization as $\left|2Q_{lkij}L_{lk}\dfrac{\mu_{\alpha\alpha}^S}{R}\right| \cong 10^7 - 10^9$ SI units; and so $\left|T_{CR}^i - T_C^i\right| \sim 50-500$ K at $L_{lk} \sim 1$ and $\alpha_T \sim 10^6$ m/FK for $|\mu_{\alpha\alpha}^S| = 5-50$ N/m. Thus, under the favorable conditions surface stress essentially increases the transition temperature and may induce ordered phase in incipient ferroelectrics.

## 3. Free energy functional and elastic problem for nanosized perovskites

For perovskite symmetry Hibbs bulk free energy expansion (1a) on polarization $P_3$ and stress $\sigma_{nm}$ powers has the form:

$$G_V = \int_V d^3 r \left( \begin{array}{c} \dfrac{a_1(T)}{2} P_3^2 + \dfrac{a_{11}}{4} P_3^4 + \dfrac{a_{111}}{6} P_3^6 + \dfrac{g}{2}(\nabla P_3)^2 \\ - P_3 \left( E_0 + \dfrac{E_3^d}{2} \right) - Q_{ij33} \sigma_{ij} P_3^2 - \dfrac{1}{2} s_{ijkl} \sigma_{ij} \sigma_{kl} \end{array} \right) \tag{5a}$$

Here $E_3^d$ and $E_0$ are depolarization and external electric field z-components.

The surface free energy (1b) polarization dependent expansion has the form

$$G_S = \sum_i \int_{S_i} d^2 r \left( \begin{array}{c} \dfrac{a_1^{Si}}{2} P_3^2 + \dfrac{a_{11}^{Si}}{4} P_3^4 + \mu_{\alpha\beta}^{Si} s_{\alpha\beta jk} \sigma_{jk} + \dfrac{v_{jklm}^{Si}}{2} \sigma_{jk} \sigma_{lm} + \\ + d_{3jk}^{Si} \sigma_{jk} P_3 - q_{jk33}^{Si} \sigma_{jk} P_3^2 \end{array} \right) \tag{5b}$$

Here superscript $Si$ numbered the surfaces; $n_k$ is the normal to the surface; Greek characters label two-dimensional indices in the surface plane, whereas Roman indices are three-dimensional ones.

For the sake of simplicity hereinafter we consider the case of mechanically isotropic solid, where the symmetry of surface stress tensors are isotropic, namely $\mu_{\alpha\beta}^{Si} = \mu \delta_{jk}$ for mechanically free nanoparticles ($\delta_{jk}$ is the Kroneker symbol).

Free energy (5) is minimal when polarization $P_3$ and relevant stress tensor components $\sigma_{jk}$ are defined at the nanostructure boundaries [24]. Under such conditions, one should solve equation of state $\dfrac{\partial G}{\partial \sigma_{jk}} = -u_{jk}$, where $u_{jk}$ is the strain tensor.

For the cases of the clamped system with defined displacement components (or with mixed boundary conditions) one should find the equilibrium state as the minimum of the Helmholtz free energy $F_V + F_S$ ($F_V = G_V + \int_V d^3 r \cdot u_{jk} \sigma_{jk}$ and $F_S = G_S + \int_S d^2 r \cdot u_\alpha \sigma_{\alpha k} n_k$) originated from Legendre transformation of $G$ [33].

Equilibrium equations of state could be obtained after variation of the Helmholtz energy on displacement $u_j$, Gibbs energy on stress $\sigma_{ij}$, polarization $P_3$ and its derivatives:

$$\dfrac{\partial \sigma_{ij}}{\partial x_i} = 0, \quad Q_{ij33} P_3^2 + s_{ijkl} \sigma_{kl} = u_{ij}, \tag{6a}$$

$$\left( a_1 - Q_{ij33} \sigma_{ij} \right) P_3 + a_{11} P_3^3 + a_{111} P_3^5 - g \dfrac{\partial^2 P_3}{\partial x_k \partial x_k} = E_0 + E_3^d. \tag{6b}$$

Eqs.(6) should be supplemented by the boundary conditions for strain (or stress) and polarization. To the best of our knowledge the general solution of the coupled problem given by Eqs.(6) is absent. In what follows we will use decoupling approximation for mechanical and electrostatic equations allowing for the boundary conditions on the nanostructure surfaces.

*(a) Freestanding cylindrical particle*

The boundary conditions (2) on the surface of cylindrical rod of radius $R$ in the cylindrical coordinates $(r, \varphi, z)$ have the following form:

$$\sigma_{\rho\rho}\big|_{\rho=R} = -\frac{\mu}{R}, \quad \sigma_{\rho\varphi}\big|_{\rho=R} = 0, \quad \sigma_{\rho z}\big|_{\rho=R} = 0,$$
$$\sigma_{zz}\big|_{z=\pm h/2} = 0, \quad \sigma_{z\rho}\big|_{z=\pm h/2} = 0, \quad \sigma_{z\varphi}\big|_{z=\pm h/2} = 0 \tag{7}$$

Equation $\partial \sigma_{ij}/\partial x_i = 0$ in the bulk of cylindrical body along with the boundary conditions (7) can be fulfilled with uniform solution. The stress and strain tensor components have the following form:

$$\sigma_{\rho\rho} = \sigma_{\varphi\varphi} = \sigma_{11} = \sigma_{22} = -\frac{\mu}{R},$$
$$\sigma_{\rho\varphi} = \sigma_{\rho z} = \sigma_{zz} = \sigma_{z\varphi} = \sigma_{12} = \sigma_{13} = \sigma_{23} = \sigma_{33} = 0, \tag{8a}$$

$$u_{11} = u_{22} = u_{\varphi\varphi} = u_{\rho\rho} = -(s_{11} + s_{12})\frac{\mu}{R} + Q_{12}P_3^2,$$
$$u_{33} = -s_{12}\frac{\mu}{R} + Q_{11}P_3^2, \quad u_{23} = u_{13} = u_{12} = 0. \tag{8b}$$

*(b) Freestanding spherical particle*

The boundary conditions (2) on the surface of spherical particle of radius $R$ have the following form in the spherical coordinates $(r, \theta, \varphi)$:

$$\sigma_{rr}\big|_{r=R} = -\frac{2\mu}{R}, \quad \sigma_{r\varphi}\big|_{r=R} = 0, \quad \sigma_{r\theta}\big|_{r=R} = 0. \tag{9}$$

Equation $\partial \sigma_{ij}/\partial x_i = 0$ in the bulk of spherical body along with the boundary conditions (9) can be fulfilled with uniform solution. The stress and strain tensor components have the form:

$$\sigma_{rr} = \sigma_{\theta\theta} = \sigma_{\varphi\varphi} = \sigma_{11} = \sigma_{22} = \sigma_{33} = -\frac{2\mu}{R},$$
$$\sigma_{r\theta} = \sigma_{r\varphi} = \sigma_{12} = \sigma_{13} = \sigma_{23} = 0. \tag{10a}$$

$$u_{11} = u_{22} = -(s_{11} + 2s_{12})\frac{2\mu}{R} + Q_{12}P_3^2, \quad u_{33} = -(s_{11} + 2s_{12})\frac{2\mu}{R} + Q_{11}P_3^2,$$
$$u_{23} = 0, \quad u_{13} = 0, \quad u_{12} = 0. \tag{10b}$$

## 4. Euler-Lagrange equation for cylindrical nanoparticle in ambient conditions

Let us consider ferroelectric cylindrical nanoparticle with radius $R$, height $h$ and axisymmetric polarization $P_3(\rho, z)$ oriented along z –axes. The external electric field is $\mathbf{E} = (0,0,E_0)$ (see Fig. 1).

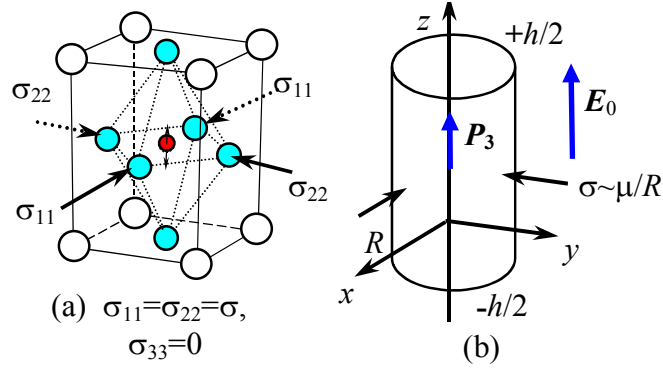

**FIG. 1. (a)** The perovskite unit cell deformation. **(b)** Geometry of cylindrical particle.

For a perovskite ferroelectric nanorod with polarization $P_3(\rho)$ the substitution of solution (8) into the free energy (5) gives the expression for the polarization-dependent part:

$$G_V = h\int_0^R r\,dr\left(\left(\frac{a_1}{2} + 2Q_{12}\frac{\mu}{R}\right)P_3^2 + \frac{a_{11}}{4}P_3^4 + \frac{a_{111}}{6}P_3^6 + \frac{g}{2}\left(\frac{\partial}{\partial\rho}P_3\right)^2 - P_3\left(E_0 + \frac{E_d}{2}\right)\right) \quad (11a)$$

$$G_S = hR\left(\left(\frac{a_1^S}{2} + 2q_{12}^S\frac{\mu}{R}\right)P_3^2(R) + \frac{a_{11}^S}{4}P_3^4(R) - 2d_{31}^S\frac{\mu}{R}P_3(R)\right) \quad (11b)$$

Hereinafter we neglect depolarization field $E_d$ for the case $h \gg R$ of the considered long nanorods (nanowires). Variation of free energy (11) leads to the Euler-Lagrange equation for the polarization $P_3(\rho)$:

$$\begin{cases}\left(a_1 + 4Q_{12}\frac{\mu}{R}\right)P_3(\rho) + a_{11}P_3^3(\rho) + a_{111}P_3^5(\rho) - g\frac{1}{\rho}\frac{\partial}{\partial\rho}\rho\frac{\partial}{\partial\rho}P_3(\rho) = E_0, \\ \left.\left(P_3 + \lambda_S\left(\frac{dP_3}{d\rho} + \frac{a_{11}^S}{g}P_3^3\right)\right)\right|_{\rho=R} = -P_d,\end{cases} \quad (12)$$

Where the boundary conditions have been rewritten via renormalized characteristic length $\lambda_S$ and surface polarization $P_d$, namely:

$$\lambda_S^{-1}(R) = \frac{a_1^S}{g} + \frac{4q_{12}^S}{g}\frac{\mu}{R}, \quad (13a)$$

$$P_d(R) = -\frac{2\mu}{R}d_{31}^S\frac{\lambda_S}{g}. \quad (13b)$$

It is worth to underline, that characteristic length $\lambda_S$ could be negative or positive, since the signs of $a_1^S$ and $q_{12}^S$ are not predetermined. When surface piezoelectric effect could be absent (i.e. $P_d = 0$), then characteristic length $\lambda_S$ had the meaning of extrapolation length. Introducing the following parameters $R_\lambda = -4q_{12}^S\mu/a_1^S$ and $\lambda_g = g/a_1^S$, one obtains that $\lambda_S^{-1}(R) = \lambda_g^{-1}(1 - R_\lambda/R)$. Normalized characteristic length $\lambda_S/|\lambda_g|$ size dependence is shown in Figs. 2. It is clear that $\lambda_S/|\lambda_g|$ is negative at $\mu > 0$ and $R < R_\lambda$ as it should be expected from Eqs.(13a).

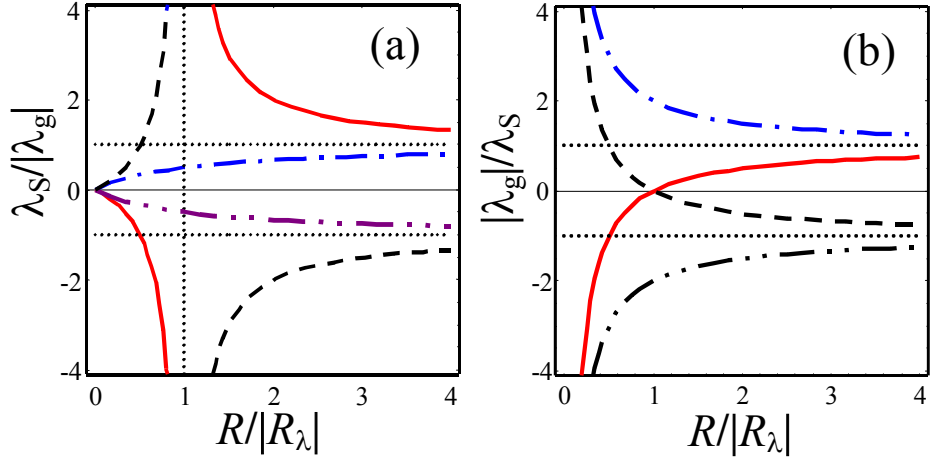

**FIG. 2.** Characteristic length $\lambda_S/|\lambda_g|$ (a) and $\lambda_S/|\lambda_g|$ (b) vs. nanowire radius $R/|R_\lambda|$ for $\lambda_g > 0$, $R_\lambda > 0$ (solid curves); $\lambda_g < 0$, $R_\lambda > 0$ (dashed curves); $\lambda_g > 0$, $R_\lambda < 0$ (dash-dotted curves) and $\lambda_g < 0$, $R_\lambda < 0$ (double-dash-dotted curves).

It should be noted that some authors (see e.g. [15], [16], [17]) also considered size-dependent characteristic (extrapolation) length that was derived from microscopic models and thus could be related with surface stress considered here in the phenomenological framework.

Application of the direct variational method for the Euler-Lagrange equation (12) approximate solution as it was described earlier [7, 34] leads to the conventional form of the free energy with renormalized coefficients. In particular surface polarization $P_d$ in the boundary conditions (12) leads to the built-in field $E_{cyl}$ appearance that induces electret-like polar state at $R < R_{cr}$ (instead of conventional paraelectric phase) and smears the dielectric permittivity maximum in the phase transition point. The electret-like state posses piezoelectric and pyroelectric properties, but hysteresis loops are absent [35]. For cylindrical nanoparticle the built-in field has the form

$$E_{cyl}(R) \approx -\frac{4\mu}{R^2}d_{31}^S. \qquad (14)$$

One can see from Eq.(14) that the built-in field is proportional to the surface stress tensor $\mu$ and piezoelectric coefficients $d_{31}^S$, it increases with radius decrease.

Transcendental equation for the determination of Curie temperature $T_{cyl}(R)$ at a given radius $R$ as well as for the critical radius $R_{cyl}(T)$ at a given temperature $T$ that corresponds to the second order phase transition from ferroelectric to paraelectric phase (at $E_{cyl} = 0$) or electret-like state (at $E_{cyl} \neq 0$) acquires the form:

$$J_0\left(R\sqrt{-\frac{a_R(T,R)}{g}}\right) - \lambda_S(R)\sqrt{-\frac{a_R(T,R)}{g}} J_1\left(R\sqrt{-\frac{a_R(T,R)}{g}}\right) = 0. \quad (15)$$

Where $J_0$ and $J_1$ are Bessel functions of the zero and first order correspondingly; $a_R(T,R) = a_1(T) + 4Q_{12}\frac{\mu}{R}$.

*4.1. Ferroelectricity enhancement in conventional ferroelectric nanorods*

For *conventional* ferroelectrics Pade approximations of Eq.(15) solution for $T_{cyl}(R)$ could be rewritten as:

$$T_{cyl}(R) \approx \begin{cases} T_C - \frac{4Q_{12}}{\alpha_T}\frac{\mu}{R} - \frac{g}{\alpha_T}\frac{2}{R\lambda_S(R) + 2R^2/k_{01}^2}, & \lambda_S(R) \geq 0, \\ T_C - \frac{4Q_{12}}{\alpha_T}\frac{\mu}{R} - \frac{g}{\alpha_T}\left(\frac{2}{R\lambda_S(R)} - \frac{1}{\lambda_S^2(R)}\right), & \lambda_S(R) < 0. \end{cases} \quad (16)$$

Where $k_{01} = 2.408...$ is the smallest positive root of equation $J_0(k) = 0$. Note, that at $\lambda_S \to 0$ Eq.(16) reduces to the one obtained in Refs. [7, 8].

Under the condition $q_{12}^S = 0$ we derived the following approximate expression for the critical radius $R_{cyl}(T)$ at a given temperature $T$:

$$R_{cyl}(T) \approx \begin{cases} -\frac{k_{01}^2}{4}\lambda_g - \frac{2Q_{12}\mu}{a_1(T)} \pm \sqrt{\left(\frac{k_{01}^2}{4}\lambda_g - \frac{2Q_{12}\mu}{a_1(T)}\right)^2 - \frac{g}{a_1(T)}\frac{k_{01}^2}{4}}, & \lambda_g \geq 0; \\ \frac{2(g\lambda_g + 2Q_{12}\mu\lambda_g^2)}{g - a_1(T)\lambda_g^2}, & \lambda_g < 0. \end{cases} \quad (17)$$

It is obvious that physically relevant values of critical radius should be positive. That is why signs $\pm$ before the radical correspond to the different conditions. Namely at $Q_{12}\mu < 0$ and $a_1(T) < 0$ or ($Q_{12}\mu > 0$ and $a_1(T) > 0$) only sign "+" makes sense, while at $Q_{12}\mu < 0$ and $a_1(T) > 0$ both signs "$\pm$" have sense and both critical radiuses define the region where ferroelectricity exists. In the case $(g + 2Q_{12}\mu\lambda_S) < 0$ and $\lambda_S \geq 0$ the region of $T_{cyl}(R) > T_C$ may extend down to $R = 0$. The simplest expression corresponds to the case $\lambda_S = 0$ (i.e. $q_{12}^S = 0$ and $a_1(T) = 0$), when

$$R_{cyl}(T) \approx -\frac{2Q_{12}\mu}{a_1(T)} \pm \sqrt{\left(\frac{2Q_{12}\mu}{a_1(T)}\right)^2 - \frac{g}{a_1(T)}\frac{k_{01}^2}{4}}.$$

Hereinafter we consider the most favorable case $\mu Q_{12} < 0$ for ferroelectricity conservation in perovskite nanowires. It is seen from Eqs.(16)-(17) that the tension radii $R_\mu = -4Q_{12}\mu/\alpha_T T_C$ and $R_\lambda = -4q_{12}^S\mu/a_1^S$, length $\lambda_g = g/a_1^S$ and bulk correlation radius at zero temperature $R_S = \sqrt{g/\alpha_T T_C}$ [20] determine the phase diagram. Note, that no restrictions are known for phenomenological parameters $R_\lambda$ and $\lambda_g$, since the quantity $a_1^S$ can take arbitrary values. Ferroelectric phase transition temperature $T_{cyl}/T_C$ vs. radius $R/R_S$ for different $R_\mu/R_S$ ratio and parameters $R_\lambda/R_S$, $\lambda_g/R_S$ determining $\lambda_S^{-1}(R) = \lambda_g^{-1}(1 - R_\lambda/R)$ dependence is depicted in Figs. 3.

The most exciting result is the transition temperature enhancement $T_{cyl} \gg T_C$ at small radius $R \ll R_S$ (see curves 4, 5 in Fig. 3). The case corresponds to the so-called surface polar state [23], appeared at negative $\lambda_S$ values. However, under the condition $a_1^S > 0$, $\lambda_S$ is positive at $R > R_\lambda$, and $\lambda_S \to g/a_1^S$ at $R \to \infty$ in accordance with Eq.(13a), as well as electrostriction term $\sim Q_{12}\mu/R$ vanishes at $R \to \infty$, making clear that no ferroelectricity enhancement appears in the bulk.

As one could expect, there is a wide range of $R/R_S$ values where $T_{cyl}/T_C < 1$ for the chosen parameters including the point $T_{cyl} = 0$ (see Fig. 3). The point corresponds to the minimal critical radius $R_{cyl}(0)$ of the size-driven ferroelectric phase transition.

Under the favorable conditions size effects (confined geometry) enhance ferroelectricity in conventional ferroelectrics. In particular for nanowires of radius $R/R_S < 2-5$ at relatively large $R_\lambda/R_S$ and $R_\mu/R_S$ values, the ratio $T_{cyl}/T_C > 1$ and it increases with $R/R_S$ decrease. Note, that the values of $R_\mu$ and $R_\lambda$ are defined by the surface stress coefficient $\mu$, bulk $Q_{12}$ and surface $q_{12}^S$ electrostriction coefficients respectively. Therefore exactly these quantities are responsible for the ferroelectricity enhancement in nanowires. The increase of $T_{cyl}$ with $\lambda_g/R_S \sim \sqrt{g}$ increase may reflect the importance of the polarization gradient contribution.

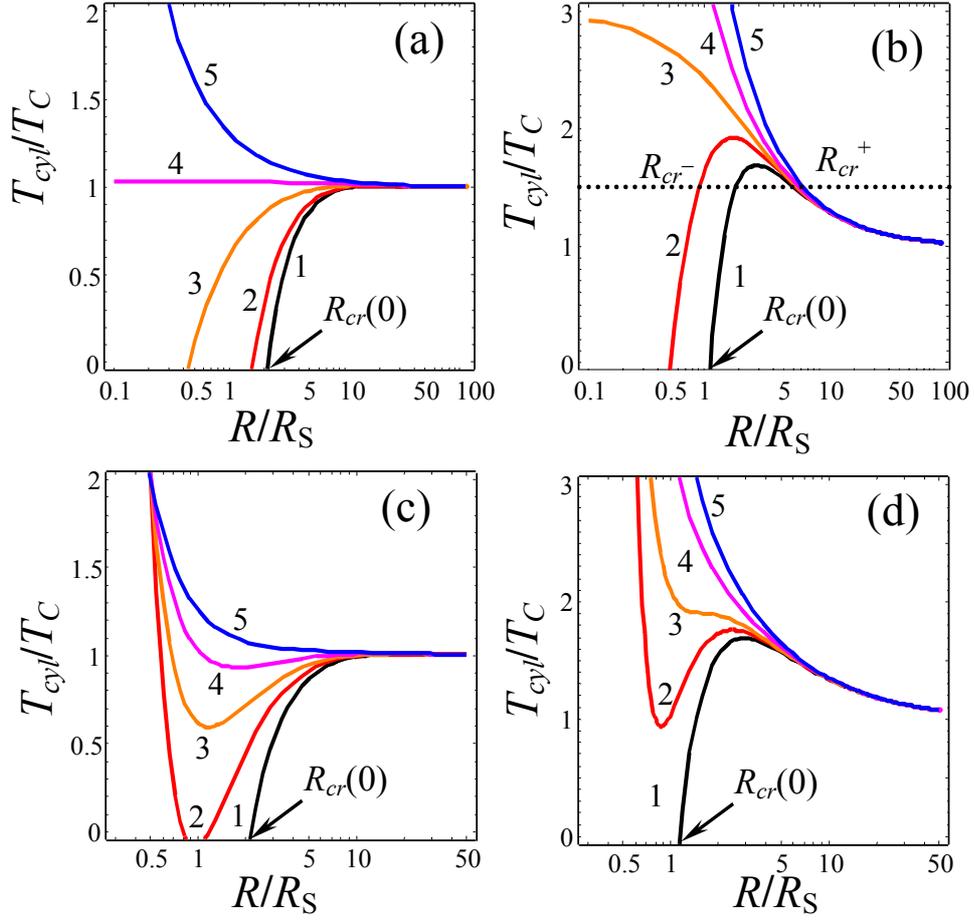

**FIG. 3.** Ferroelectric phase transition temperature $T_{cyl}/T_C$ vs. $R/R_S$ for **(a)** $R_\lambda/R_S = 0$, $R_\mu/R_S = 0.5$, $\lambda_g/R_S = 0, 0.5, 2, 4, 10$, (curves 1, 2, 3, 4, 5); **(b)** $R_\lambda/R_S = 0$, $R_\mu/R_S = 4$, $\lambda_g/R_S = 0, 0.3, 0.5, 1, 4$ (curves 1, 2, 3, 4, 5); **(c)** $R_\lambda/R_S = 0.5$, $R_\mu/R_S = 0.5$, $\lambda_g/R_S = 0, 0.5, 1, 2, 4$ (curves 1, 2, 3, 4, 5); **(d)** $R_\lambda/R_S = 0.5$, $R_\mu/R_S = 4$, $\lambda_g/R_S = 0, 0.1, 0.2, 0.5, 1$ (curves 1, 2, 3, 4, 5).

It is clear that size effects are most pronounced in the region $R < 10 R_S$ that typically corresponds to the nanowires radius less than 50nm since $R_S \sim 5-50\,\text{A}^\circ$; at that the region width and the ratio $T_{cyl}/T_C$ increases with the values of $R_\mu$ and/or $R_\lambda$ increase, the latter parameters reflect the contribution of the surface stress.

As it has been already mentioned in the introduction, the most striking observation of ferroelectricity enhancement and conservation in long nanorods (radius 15nm, length 500nm) has been reported by Yadlovker and Berger [6]. Besides ferroelectricity conservation up to Roshelle salt decomposition temperature that was explained by us earlier [7], [8], the authors obtained the temperature dependence of remnant polarization $P_{SV}(T)$, hysteresis loops and switching time. These polar properties can be calculated by a conventional way on the basis of the free energy with renormalized coefficients:

$$G \approx \left( \alpha_T (T - T_{cyl}(R)) \frac{P_{3V}^2}{2} + a_{11} \frac{P_{3V}^4}{4} + a_{111} \frac{P_{3V}^6}{6} - P_{3V}(E_0 + E_{cyl}(R)) \right) \quad (18)$$

Under the condition of negligibly small surface piezoeffect, it is easy to obtain from (18) the spontaneous polarization is $P_{SV} \approx \sqrt{\alpha_T (T_{cyl}(R) - T)/a_{11}}$ and thermodynamic coercive field is $E_C^T = \frac{2P_{SV}}{3\sqrt{3}} \alpha_T (T - T_{cyl}(R))$. The thermodynamic coercive field corresponds to the case of nanorod homogeneous (mono-domain) switching. In the case of inhomogeneous switching process the activation field $E_{cr}^a$ necessary for domain nucleus onset determines the observed coercive field $E_C$ and switching time [20]. For the latter case renormalized coefficient $\alpha_T (T - T_{cyl}(R))$ should be included into the domain wall energy, spontaneous polarization, dielectric permittivity, *etc* instead of the bulk coefficient $a_1(T)$, in order to describe adequately the domain nucleation stage and further domain wall motion (see Appendix A for details).

The applicability of our model to the description of phase transition between cubic paraelectric and tetragonal ferroelectric phases in ferroelastic – ferroelectric RS is valid only in the temperature range where RS ferroelectric properties can be described by the phenomenological framework. Moreover, we neglected the piezoelectric effect with respect to the shear stress in the paraelectric phase of RS since the effective surface tension creates no tangential stresses.

Despite the aforementioned limitations, let us perform quantitative comparison with experimental data for RS. The comparison of experimentally obtained in Ref.[6] polarization temperature dependence $P_{SV}(T)$ and hysteresis loop $P_{SV}(E_0)$ (symbols) with theoretical calculations (solid curves) are presented in Figs.4.

The dependence $P_{SV}(T)$ was calculated from the equation $\alpha_T (T - T_{cyl}(R))P_{3V} + a_{11}P_{3V}^3 \approx E_{cyl}(R)$ for RS material parameters (see Fig.4a).

Calculated in Appendix A activation field value $E_{cr}^a \approx 13\,\text{kV/cm}$ is very close to the experimentally obtained coercive field $E_C \approx 13.6\,\text{kV/cm}$ [6]. This means that thermal fluctuations ~ $k_B T$ cause rapid nanodomain nucleation in the nanorod under applied field $E_{cr}^a \approx 13\,\text{kV/cm}$, in contrast to the switching of bulk sample with much smaller coercive fields about 0.2 kV/cm. Using the value $P_{SV} \approx 0.28\,\mu\text{C/cm}^2$ calculated at $T = 21\,°\text{C}$ as a remnant polarization (see Fig.4a) and the value $E_{cr}^a \approx 13\,\text{kV/cm}$ as a coercive field we modeled hysteresis loop $P_{SV}(E_0)$ from a conventional kinetic equation [20] (see Fig.4b).

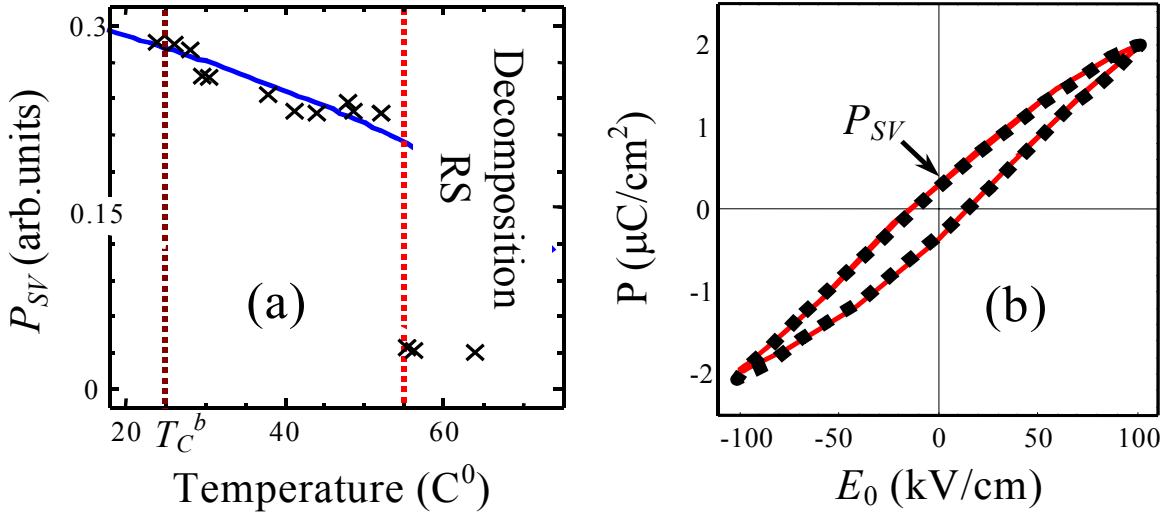

**FIG.4**. The dependence of remnant polarization $P_{SV}(T)$ on temperature (a) and hysteresis loop (b) of RS nanorods with radius $R = 15$ nm. Symbols are experimental data from Ref. [6]: (a) remnant polarization at applied field amplitude $E_0^{max} = 30$ kV/cm; (b) hysteresis loop at applied field frequency 1kHz and $T = 21$°C. Solid curves are theoretical calculations with parameters $T_{cyl} = 80$°C, $|E_{cyl}| \approx 2$ kV/cm (a) and $P_{SV} \approx 0.28\,\mu\text{C/cm}^2$, $E_C \approx 13$ kV/cm (b).

Using RS material parameters $g \approx 9 \cdot 10^{-11}\,\text{m}^3/\text{F}$ [36], $\alpha_T \approx 7.74 \cdot 10^7$ m/F·K, $T_C = 297$ K, $a_{11} = 3.36 \cdot 10^{15}$ SI units (estimated from bulk polarization of RS crystal at room temperature $P_S = \sqrt{\alpha_T T_r / a_{11}} \approx 0.25\,\mu\text{C/cm}^2$) and the sum $Q_{12} + Q_{13} = -0.63\,\text{m}^4/\text{C}^2$ (instead of $2Q_{12}$ for perovskites), our fitting values $T_{cyl} = 80$°C, $R = 15$ nm and Eq.(16), we obtained the estimation for surface stress coefficient $\mu \approx 25$ N/m. Note, that it is the upper estimation (i.e. $\mu \leq 25$ N/m) since the third term in Eq.(16) appeared either positive at negative values $\lambda_S$ or negligibly small at $\lambda_S \geq 0$ allowing for the small polarization gradient coefficient value *g*. Unfortunately we could not find any experimental data concerning surface piezoeffect $d_{ij}^S$ in RS, so we regard built-in field $E_{cyl} \sim d_{31}^S$ as a fitting parameter and obtained that $|E_{cyl}| \leq 2$ kV/cm and thus $d_{31}^S \approx 7.2 \cdot 10^{-16}$ m$^2$/V. It is seen that theory fitted experimental data rather well at reasonable values of surface stress.

It is worth to mention that some other experimental results indirectly speak in favor of the ferroelectricity enhancement and conservation in PbZr$_{0.52}$Ti$_{0.48}$O$_3$ nanorods with diameter less than 10-20nm [37], [38] and in single-crystalline PbZr$_{0.2}$Ti$_{0.8}$O$_3$ nanowires of elliptic cross section with main diameters of 70 and 180 nm [32].

The developed phenomenological approach describes the observed peculiarities of ferroelectric nanorods.

*4.2. Size-induced ferroelectricity in incipient ferroelectric nanowires*

The Barrett formulae $a_1(T) = \alpha_T \left( \frac{T_q}{2} \coth\left(\frac{T_q}{2T}\right) - T_0 \right)$ is valid for both *incipient* and *conventional* ferroelectrics [39] at wide temperature interval including low (quantum) temperatures. At temperatures $T \gg T_q/2$ Barrett formulae transforms into the classical form $a_1(T \gg T_q) \approx \alpha_T(T - T_0)$.

The transition temperature (induced by surface and size effects) is given by:

$$T_{cyl}(R) \approx \begin{cases} \frac{T_q}{2}\left(\text{arccoth}\left(\frac{2}{T_q}\left(T_0 - \frac{4Q_{12}}{\alpha_T}\frac{\mu}{R} - \frac{g}{\alpha_T}\frac{2}{R\lambda_S(R) + 2R^2/k_{01}^2}\right)\right)\right)^{-1}, & \lambda_S(R) \geq 0, \\ \frac{T_q}{2}\left(\text{arccoth}\left(\frac{2}{T_q}\left(T_0 - \frac{4Q_{12}}{\alpha_T}\frac{\mu}{R} - \frac{g}{\alpha_T}\left(\frac{2}{R\lambda_S(R)} - \frac{1}{\lambda_S^2(R)}\right)\right)\right)\right)^{-1}, & \lambda_S(R) < 0. \end{cases} \quad (19)$$

Using the asymptotic relation $(\text{arccoth}(x))^{-1} \to x$ for $x \gg 1$, one can obtain that at temperatures $T \gg T_q/2$ Eq.(19) tends to Eq.(16) after substitution $T_C \to T_0$, where $T_0 \leq 0$ is possible.

It is worth to underline that the second term in brackets of Eq. (19) represents the contribution of biaxial stress originated from the intrinsic surface stress.

Note that approximate expressions (17) for the critical radius $R_{cyl}(T)$ are valid after substitution of the Barrett formulae for $a_1(T)$.

Let us introduce tension radius $R_\mu = -4Q_{12}\mu/\alpha_T|T_0|$, bulk correlation radius $R_S = \sqrt{g/\alpha_T|T_0|}$ and the ratio $T_q/2T_0$. The dependence of the temperatures $T_{cyl}/|T_0|$ vs. radius $R/R_S$ is depicted in Figs. 5.

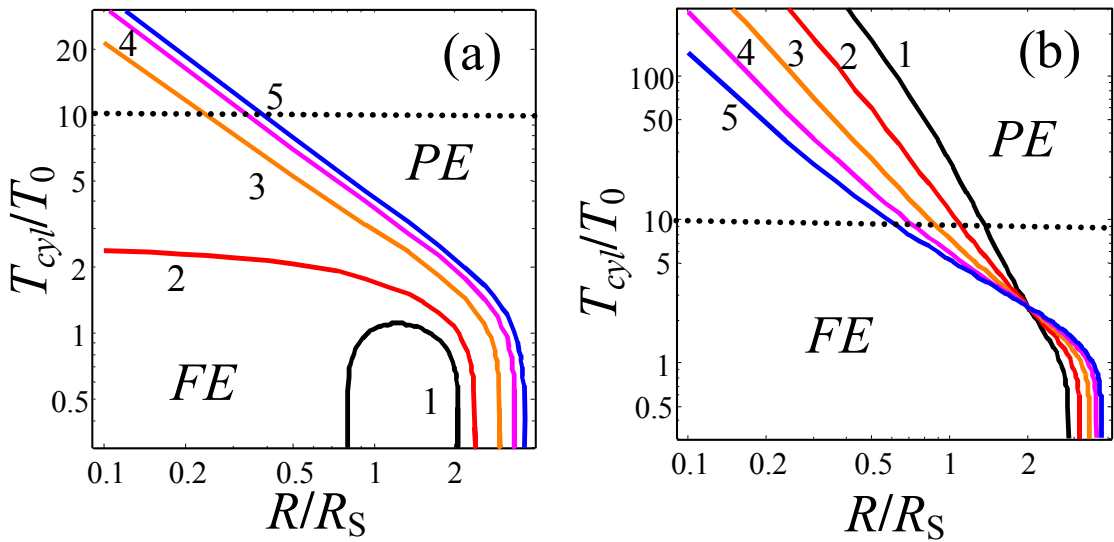

**FIG. 5.** Ferroelectric phase transition temperature $T_{cyl}/|T_0|$ vs. $R/R_S$ for incipient ferroelectric at $\lambda_S^{-1}(R) = \lambda_g^{-1}(1 - R_\lambda/R)$, $T_q/2T_0 = 2$ and **(a)** $R_\mu/R_S = 4$, $R_\lambda/R_S = 0$, $\lambda_g/R_S = 0.4; 0.5; 1; 2; 4$ (curves 1, 2, 3, 4, 5); **(b)** $R_\lambda/R_S = 2$, $\lambda_g/R_S = 0.25; 0.5; 1; 2; 4$ (curves 1, 2, 3, 4, 5).

It is clear that under the favorable conditions size effects induces ferroelectric phase in incipient ferroelectrics. In particular in nanorods with small enough ratio $R/R_S < 0.5$ and relatively large $R_\mu/R_S$ ratio, the ratio $T_{cyl}/|T_0| \gg 1$; also the transition temperature essentially increases with $R/R_S$ decrease. The cross-point of the curves in plot (b) corresponds to the point $R = R_\lambda$, where all lengths $\lambda_S$ diverge in accordance with Eqs.(13a) and so the transition temperature becomes equal.

Usually $T_0$ is small enough and about 10-50K. However under the condition $T_{cyl}/|T_0| > 10$ (see horizontal lines in Figs.5) size-induced ferroelectricity may exist at room temperatures. Allowing for small values of $T_0$ for incipient ferroelectric and large values of correlation radius $R_S$ (from several to tens of lattice constant), the typical range $0 < R/R_S < 1$ (where $T_{cyl}/|T_0| > 0$) may be rather wide: from dozens to hundreds of lattice constants.

It is worth to underline, that under the condition $\mu Q_{12} < 0$ bi- and uni- axial radial stress should increase the transition temperature in incipient ferroelectrics. Really, Uwe and Sakudo [40] have found that the uniaxial stress higher than $5.25 \times 10^8$ Pa induces ferroelectric phase transition in bulk KTaO$_3$ at temperature 2K. The same radial stress $\sigma = \mu/R$ appeared in KTaO$_3$ nanowires of radius $R = 4 - 40$ nm at the surface stress values $\mu = 4 - 40$ N/m [29], [31] reasonable for perovskites. This means that surface stress existing under the curved surface of KTaO$_3$ nanorod could induce ferroelectricity.

Let us consider the appearance of the size-induced ferroelectric phase in KTaO$_3$ nanowires quantitatively. We used Eqs.(19) for the transition temperature; electrostriction constant $Q_{12}$ were evaluated from the experiments [40]. Ferroelectric phase transition temperature $T_{cyl}$ vs. nanowire radius $R$ for KTaO$_3$ is shown in Fig.6.

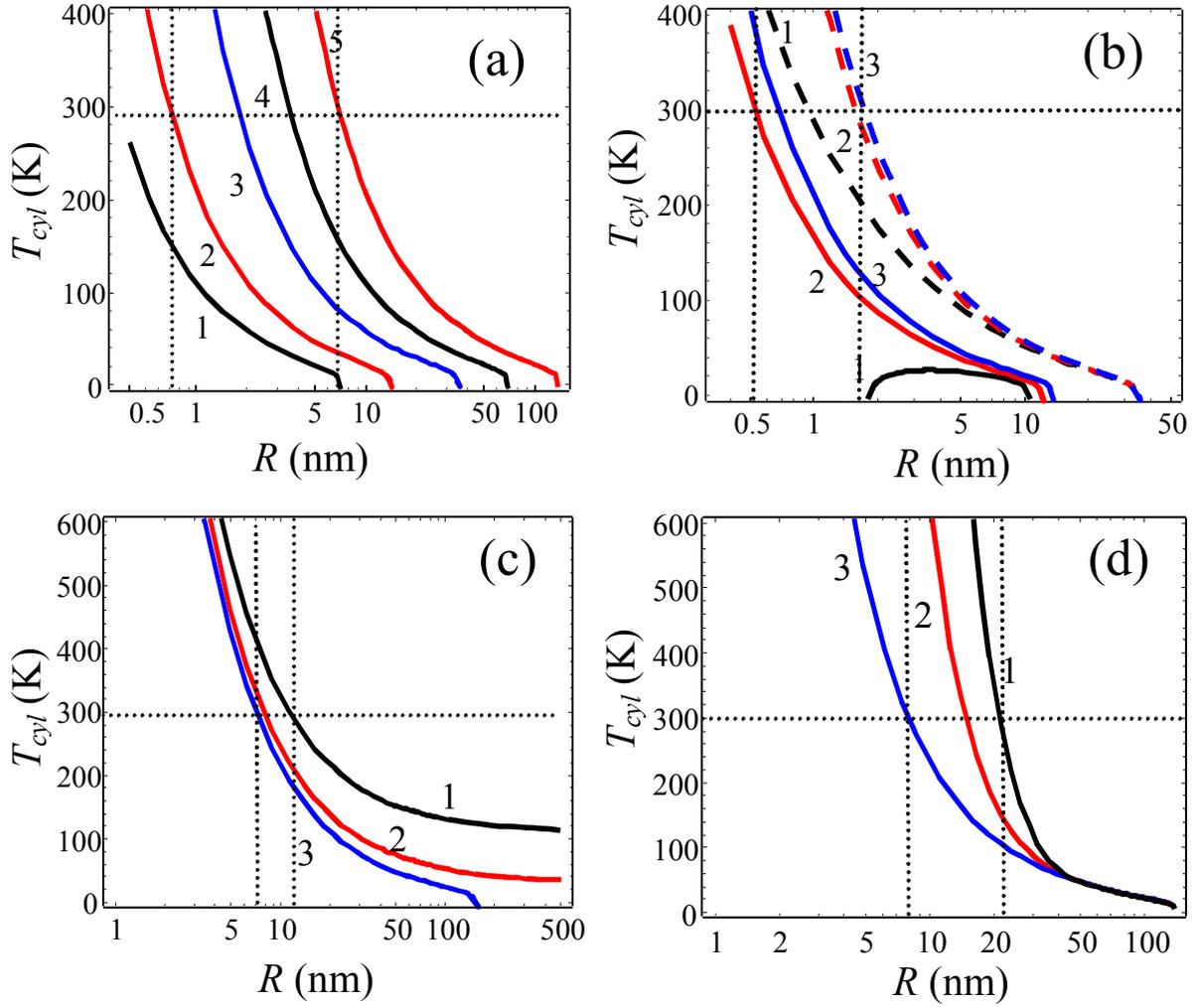

**FIG. 6.** Ferroelectric phase transition temperature $T_{cyl}$ vs. nanowires radius $R$ for KTaO$_3$ material parameters $T_q = 55$ K, $T_0 = 13$ K, Curie-Weiss constant $C_{CW} = 5.6 \times 10^4$ K, $Q_{12} = -0.023$ m$^4$/C$^2$; gradient coefficient $g = 10^{-10}$ V m$^3$/C in SI and $\lambda_S^{-1}(R) = \lambda_g^{-1}(1 - R_\lambda/R)$. **(a)** $\lambda_g \to +\infty$, $R_\lambda = 0$ and different surface stress values $\mu$: 2; 4; 10; 20; 40 N/m (curves 1, 2, 3, 4, 5); **(b)** $\mu = 4$ N/m (solid curves) and $\mu = 10$ N/m (dashed curves), $R_\lambda = 0$ and $\lambda_g = 0.6, 4, 40$ nm (curves 1, 2, 3); **(c)** $\mu = 40$ N/m, $R_\lambda = 0$ and $\lambda_g = -1; -2; -10$ nm (curves 1, 2, 3); **(d)** $\mu = 40$ N/m, $R_\lambda = 50$ nm and $\lambda_g = 1, 2, 10$ nm (curves 1, 2, 3).

Part (a) corresponds to the case when characteristic length $\lambda_S \to +\infty$ (i.e. $q_{12}^S = 0$ and $\lambda_g \to +\infty$) and so polarization gradient can be neglected, and radius dependence of $T_{cyl}$ is caused by the surface stress only. Part (b) and (c) correspond to the case when both surface stress and polarization gradient contribute into the transition temperature, but surface electrostriction is absent, i.e. $q_{12}^S = 0$ and so $\lambda_S = \lambda_g = const$. It is clear that negative $\lambda_S$ increases the transition temperature in comparison with the positive ones (compare parts (b) and (c)). Part (d) shows the influence of negative surface electrostriction on $T_{cyl}$. The cross-point of the curves in plot (d) corresponds to the point $R = R_\lambda$, where all characteristic lengths diverges in accordance with Eqs.(13a).

The prediction of size-induced ferroelectricity in KTaO$_3$ nanorods of radius less then 5-20 nm (see vertical lines in Fig.6) at room temperatures (see horizontal lines in Fig.6) could be very important for applications. Since $Q_{12} < 0$, the effect is possible for positive intrinsic surface stress coefficient μ that compresses the particle. Additional desirable condition is the negative length $\lambda_S$, possible even at $\lambda_g > 0$ for nanoparticle radius $R < R_\lambda$ when $\mu q_{12}^S < 0$.

Thus, we came to the conclusion about size-induced ferroelectricity in incipient ferroelectric KTaO$_3$ at room temperature for small enough nanowires. It is obvious that the similar size-induced transition could be found in another incipient ferroelectric SrTiO$_3$.

## 5. Euler-Lagrange equation for spherical nanoparticle in ambient conditions

Let us consider ferroelectric perovskite spherical nanoparticle of radius $R$ and polarization $P_3(r)$ oriented along z –axes. The external electric field is $\mathbf{E} = (0, 0, E_0)$ (see Fig. 7).

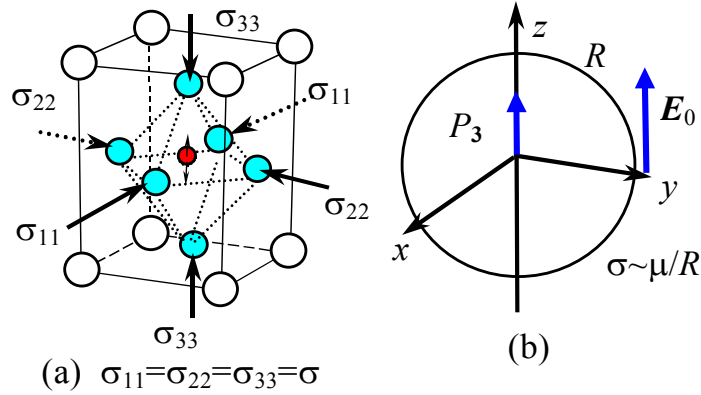

**FIG. 7 (a)** The unit cell deformation under the hydrostatic pressure $\sigma_{jj} = -2\mu/R$. **(b)** Geometry of spherical particle.

For sphere with polarization $P_3(r)$ substitution of solution (8) into the free energy (5) leads to the following expression for the polarization-dependent part:

$$G_V = \int_0^R r^2 dr \left( \left( \frac{a_1}{2} + (Q_{11} + 2Q_{12})\frac{2\mu}{R} \right) P_3^2 + \frac{a_{11}}{4} P_3^4 + \frac{a_{111}}{6} P_3^6 + \frac{g}{2}\left( \frac{\partial}{\partial r} P_3 \right)^2 - P_3 \left( E_0 + \frac{E_3^d}{2} \right) \right) \quad (20a)$$

$$G_S = R^2 \left( \left( \frac{a_1^S}{2} + (2q_{12}^S + q_{11}^S)\frac{2\mu}{R} \right) P_3^2(R) + \frac{a_{11}^S}{2} P_3^4(R) - (2d_{31}^S + d_{33}^S)\frac{2\mu}{R} P_3(R) \right) \quad (20b)$$

Here depolarization field is $E_3^d = n_d(\overline{P_3} - P_3)$, where $\overline{P_3}$ stands for the spatial average on the sample volume, $n_d = \dfrac{4\pi}{1+2\varepsilon_e}$ is a depolarization factor, $\varepsilon_e$ is the nanoparticle ambient dielectric permittivity [41]. Variation of free energy (20) leads to the Euler-Lagrange equation for polarization $P_3(r)$:

$$\begin{cases} \left(a_1 + (Q_{11} + 2Q_{12})\dfrac{4\mu}{R}\right)P_3(r) + a_{11}P_3^3(r) + a_{111}P_3^5(r) - \dfrac{g}{r^2}\dfrac{\partial}{\partial r}r^2\dfrac{\partial}{\partial r}P_3(r) = E_0 + E_3^d, \\ \left(P_3 + \lambda_S\left(\dfrac{dP_3}{dr} + \dfrac{a_{11}^S}{g}P_3^3\right)\right)\bigg|_{r=R} = -P_d, \end{cases} \quad (21)$$

Here the boundary conditions have been rewritten via renormalized length $\lambda_S$ and surface polarization $P_d$, namely:

$$\lambda_S^{-1}(R) = \dfrac{a_1^S}{g} + \dfrac{(2q_{12}^S + q_{11}^S)}{g}\dfrac{4\mu}{R}, \quad (22\text{a})$$

$$P_d(R) = -(2d_{31}^S + d_{33}^S)\dfrac{2\mu}{R}\dfrac{\lambda_S}{g}. \quad (22\text{b})$$

In general case length $\lambda_S$ could be negative or positive because both signs of $a_1^S$ could encounter. Introducing the following parameters $R_\lambda = -4(2q_{12}^S + q_{11}^S)\mu/a_1^S$ and $\lambda_g = g/a_1^S$, one obtains that $\lambda_S^{-1}(R) = \lambda_g^{-1}(1 - R_\lambda/R)$. Normalized characteristic length $\lambda_S/|\lambda_g|$ size dependence is shown in Figs.8.

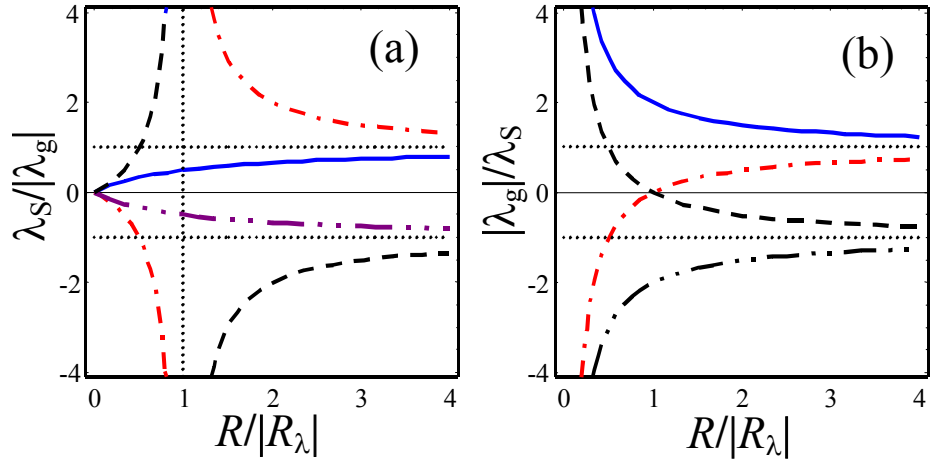

**FIG. 8.** Characteristic length $\lambda_S/|\lambda_g|$ (a) and $|\lambda_g|/\lambda_S$ (b) vs. sphere radius $R/|R_\lambda|$ for $\lambda_g > 0$, $R_\lambda < 0$ (solid curves); $\lambda_g < 0$, $R_\lambda > 0$ (dashed curves); $\lambda_g > 0$, $R_\lambda > 0$ (dash-dotted curves) and $\lambda_g < 0$, $R_\lambda < 0$ (double-dash-dotted curves).

Application of the direct variational method for Euler-Lagrange equation (21) solution leads to the conventional form of the free energy with renormalized coefficients:

$$G \approx \left( \alpha_T (T - T_{sph}(R)) \frac{P_{3V}^2}{2} + a_{11} \frac{P_{3V}^4}{4} + a_{111} \frac{P_{3V}^6}{6} - P_{3V} (E_0 + E_{sph}(R)) \right) \quad (23)$$

Here the surface polarization $P_d$ in the boundary conditions leads to the built-in electric field $E_{sph}$ appearance that induces electret-like polar state at $R < R_{cr}$ and smears the phase transition point:

$$E_{sph}(R) \approx -\frac{6\mu}{R^2} (2d_{31}^S + d_{33}^S). \quad (24)$$

One can see from Eq.(24) that the built-in field is proportional to the surface stress tensor $\mu$ and piezoelectric coefficients $d_{3j}^S$. Internal electric field $E_{sph}(R) \sim 1/R^2$ increases with particle radius decrease and can lead to the particle self-polarization when reaches thermodynamics coercive field, like it was predicted for the thin films due to the misfit strain [35]. Thus assumption about the particle single-domain state used in our consideration looks self-consistent for sizes below 100 nm in full agreement with experimental results [10], [11].

The transcendental equation for the Curie temperature $T_{sph}(R)$ at a given radius $R$ as well as for the critical radius $R_{sph}(T)$ at a given temperature $T$ that corresponds to the second order phase transition from ferroelectric to paraelectric phase (at $E_{sph} = 0$) or electret-like state (at $E_{sph} \neq 0$) acquires the form:

$$a_R + \frac{\frac{3g n_d}{R^2 (n_d + a_R)} \left( R\sqrt{\frac{n_d + a_R}{g}} \cosh\left(R\sqrt{\frac{n_d + a_R}{g}}\right) - \sinh\left(R\sqrt{\frac{n_d + a_R}{g}}\right) \right)}{\lambda_S \sqrt{\frac{n_d + a_R}{g}} \cosh\left(R\sqrt{\frac{n_d + a_R}{g}}\right) + \left(1 - \frac{\lambda_S}{R}\right) \sinh\left(R\sqrt{\frac{n_d + a_R}{g}}\right)} = 0, \quad (25)$$

where $a_R(T, R) = a_1(T) + (Q_{11} + 2Q_{12}) \frac{4\mu}{R}$. Note, that inequality $(Q_{11} + 2Q_{12}) > 0$ holds for perovskites.

*5.1. Phase transition in conventional ferroelectric nanospheres*

Under the typical condition $\varepsilon_e < 10$ and $a_1(T) = \alpha_T(T - T_C)$, transition temperature of conventional ferroelectric has the form:

$$T_{sph}(R) = T_C - (Q_{11} + 2Q_{12}) \frac{4\mu}{\alpha_T R} - \frac{3g}{R^2 \alpha_T} \theta(\lambda_S, R) \quad , \quad (26a)$$

$$\theta(\lambda_S, R) = \frac{R\sqrt{n_d/g} \cosh(R\sqrt{n_d/g}) - \sinh(R\sqrt{n_d/g})}{\lambda_S \sqrt{n_d/g} \cosh(R\sqrt{n_d/g}) + (1 - \lambda_S/R) \sinh(R\sqrt{n_d/g})}. \quad (26b)$$

At a given temperature $T$ the sphere critical radius $R_{cr}(T)$ should be found from the condition $T_{sph}(R_{cr}) = T$.

For more detailed analyses of Eqs.(26) one should take into account that the length $\lambda_S^{-1}(R) = \lambda_g^{-1}(1 - R_\lambda/R)$ depends on sphere radius $R$, parameters $R_\lambda = -(8q_{12}^S + 4q_{11}^S)\mu/a_1^S$ and $\lambda_g = g/a_1^S$. Let us introduce the radius $R_\mu = (4Q_{11} + 8Q_{12})\mu/\alpha_T T_C$, related with intrinsic surface stress, characteristic radius $R_d = \sqrt{g/n_d}$, correlation radius $R_S = \sqrt{g/\alpha_T T_C}$ that coincides with order parameter correlation radius at zero temperature [20].

Ferroelectric phase transition temperature $T_{sph}/T_C$ vs. radius $R/R_S$ calculated from Eq.(26) for different $R_\mu/R_S$ ratio and parameters $R_\lambda/R_S$, $\lambda_g/R_S$ determining $\lambda_S^{-1}(R) = \lambda_g^{-1}(1 - R_\lambda/R)$ radius dependence is depicted in Figs. 9.

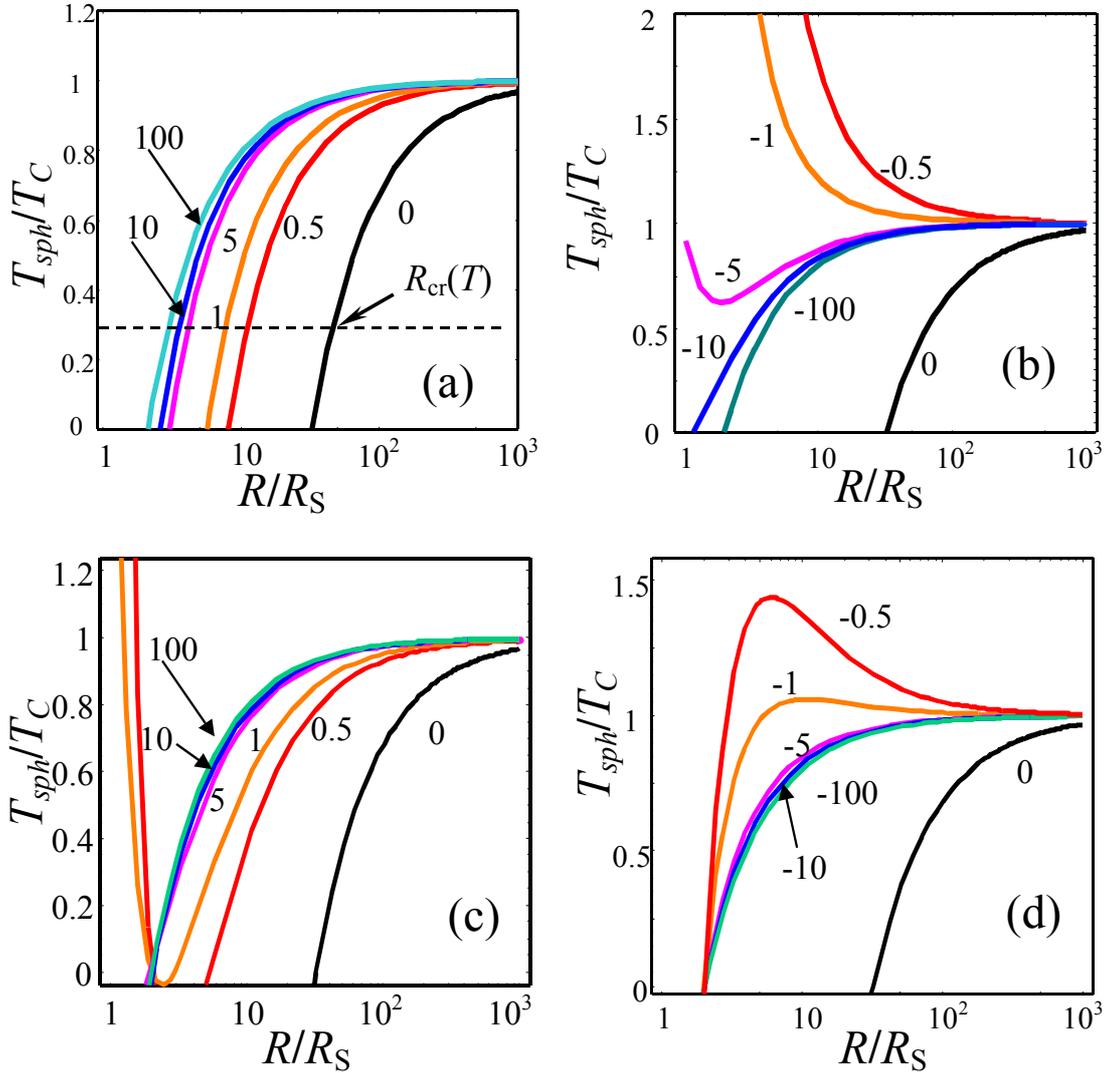

**FIG. 9.** Transition temperature $T_{sph}/T_C$ vs. $R/R_S$ calculated from (26) for conventional ferroelectric at $R_d/R_S = 0.1$, $R_\mu/R_S = 2$, $R_\lambda/R_S = -2$ **(a, b)** and $R_\lambda/R_S = +2$ **(c, d)** for positive $\lambda_g/R_S = 0$; 0.5; 1; 10; 100 **(a,c)** and negative $\lambda_g/R_S = 0$; -0.5; -1; -5; -10; -100 **(b,d)** (figures near the curves).

At positive $R_\mu$ (that corresponds to the positive surface stress coefficient $\mu > 0$, since $(Q_{11} + 2Q_{12}) > 0$) the transition temperature increase $T_{sph} > T_C$ is possible at some region of negative lengths $\lambda_g$ ($a_1^S < 0$) and/or positive $R_\lambda$ (compare with enhancement for a cylinder). The situation $\mu < 0$ might be possible at some special ambient conditions.

It is clear from Fig.9 that the critical radius $R_{cr}(0)$ belongs to the region $3R_S < R_{cr}(0) < 30 R_S$ at positive $\lambda_g/R_S$, whereas for nanowires the critical radius is much smaller at the same material parameters: $0.3 R_S < R_{cr}(0) < 3 R_S$ (if any) (compare Figs. 3a and 9a). The difference is related with the absence of depolarization field in nanowires and its presence in nanospheres.

Using the values $\mu = 0.5 - 50$ N/m [29], [31] $g = 10^{-11} - 10^{-10}$ Vm$^3$/C [42], [43], $2Q_{12} + Q_{11} = 0.03$ m$^4$/C$^2$, $\alpha_T T_C \sim 4\cdot 10^7 - 2\cdot 10^8$ in SI units, $T_C \sim 400 - 700$ K and $n_d \sim 4 - 0.04$, one can obtain for the perovskites like BaTiO$_3$ and PbTiO$_3$, that $R_\mu \approx 1.5 - 150$ nm for BaTiO$_3$ and $R_\mu \approx 0.3 - 30$ nm for PbTiO$_3$, $R_S \approx 0.4 - 1.6$ nm and $R_d \sim 0.1 - 1$ nm. Similarly to the case of nanowires, no restrictions are known for phenomenological parameters $R_\lambda$ and $\lambda_g$, since the quantity $a_1^S$ can take arbitrary values. Since typically $\exp(-R/R_d) \ll 1$ for nanospheres of radius $R > 0.5 - 5$ nm, we derived from Eqs.(26) that

$$T_{sph}(R) \approx T_C \left(1 - \frac{R_Q}{R} - \frac{R_q^2}{R^2}\right), \quad (27a)$$

$$R_{cr}(T) \approx \frac{R_Q \pm \sqrt{R_Q^2 + 4(1 - T/T_C)R_q^2}}{2(1 - T/T_C)}, \quad (27b)$$

where $R_Q = (R_\mu + 3R_S^2/\lambda_g)$ is determined by the intrinsic surface stress and bulk electrostriction and $R_q^2 = -3R_S^2 R_\lambda/\lambda_g$ is determined by the intrinsic surface stress, surface electrostriction and correlation radius $R_S$ (see Appendix A).

Comparison of typical experimental data for the dependence of Curie temperature $T_{sph}(d)$ on the size $d$ of BaTiO$_3$ and PbTiO$_3$ nanoparticles with theoretical calculations on the basis of expression (26a) as well as fit with empirical Ishikava formula $T_{cr}(R) \approx T_C\left(1 - \frac{R_0}{R - R_1}\right)$ [44] is shown in Fig.10.

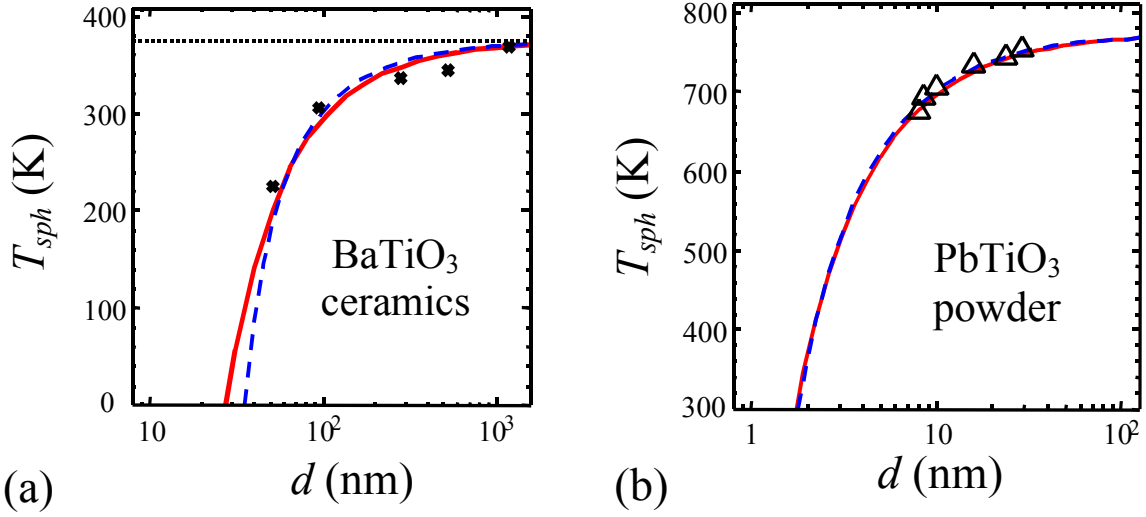

**FIG. 10**. The dependence of the Curie temperature on the mean size of the particles $d$ for dense fine grained ceramics of BaTiO$_3$ [10] (a) and nanopowder of PbTiO$_3$ obtained by EPR measurements [11] (b) Symbolas are experimental data [10] and [11], solid and dashed curves represent the fitting with Eq. (27a) and Ishikawa formula respectively. (a): $T_C = 375\,K$ and $R_Q = 19.6$ nm, $R_q = 15.5$ nm for Eq. (27a); $R_0 = 19$ nm, $R_1 = 16$ nm for Ishikawa fit. (b): $T_C = 773\,K$ and $R_Q = 1$ nm, $R_q = 0.4$ nm for Eq. (27a); $R_0 = 0.9$ nm, $R_1 = 0.3$ nm for Ishikawa fit.

It is seen from Fig. 10, that derived expression (27a) for $T_{sph}(R)$ fits the experimental points as well as purely empirical Ishikava formulae at the same amount of fitting parameters.

It is worth to note, that the critical sizes in ceramics and powder samples can vary significantly. It can be related with the ceramics preparation features, as well as with different mechanical and electrical boundary conditions for the grains of ceramics and particles of powder. In the framework of the proposed phenomenological theory the values of surface intrinsic stress $\mu$, surface energy expansion coefficient $a_1^S$, surface electrostriction $q_{ij}^S$ and depolarization factor $n_d \sim (1+2\varepsilon_e)^{-1}$ should differ for the ceramics and powder samples, prepared by different methods. Also in order to consider dielectric properties of the nanoparticles assemble, one has to take concrete expression for their size distribution function [41].

### 5.2. Size-induced ferroelectricity in incipient ferroelectric nanospheres

Using the Barrett formulae for $a_1(T)$, the transition temperature induced by surface and size effects is given by:

$$T_{sph}(R) = \frac{T_q}{2} arc\coth^{-1}\left(\frac{2}{T_q}\left(T_0 - (Q_{11} + 2Q_{12})\frac{4\mu}{\alpha_T R} - \frac{3g}{R^2\alpha_T}\theta(\lambda_S, R)\right)\right) \qquad (28)$$

Here $\theta(\lambda_S, R)$ is given by Eq.(26b). Using the asymptotic relation $(\text{arccoth}(x))^{-1} \to x$ for $x \gg 1$, one can obtain that at temperatures $T \gg T_q/2$ Eq.(28) tends to Eq.(26) after substitution $T_C^b \to T_0$.

Below we consider size-induced ferroelectric phase in KTaO$_3$ nanospheres. We used Eq.(28) for the transition temperature; electrostriction constant $(Q_{11} + 2Q_{12})$ were taken from Ref. [40]. Ferroelectric phase transition temperature $T_{sph}$ vs. sphere radius $R$ for KTaO$_3$ is shown in Fig.11. Part (a) corresponds to the case when characteristic length $|\lambda_S| \to +\infty$ ($\lambda_g \to +\infty$ and $q_{11}^S + 2q_{12}^S = 0$), so polarization gradient can be neglected and radius dependence of $T_{sph}$ is caused by the surface stress via bulk electrostriction effect. Parts (b) and (c) correspond to the case when both bulk electrostriction and polarization gradient contribute into the transition temperature, but surface electrostriction is absent, i.e. $q_{ij}^S = 0$ and so $\lambda_S = \lambda_g = const$. It is clear that negative $\lambda_g$ increases the transition temperature in comparison with the positive ones (compare (b) and (c)). Part (d) shows the influence of surface electrostriction ($q_{ij}^S \neq 0$) on transition temperature $T_{sph}$. The cross-point of the curves in plot (d) corresponds to the point $R = -R_\lambda$, where all characteristic length diverges in accordance with equation $\lambda_S^{-1}(R) = \lambda_g^{-1}(1 - R_\lambda/R)$.

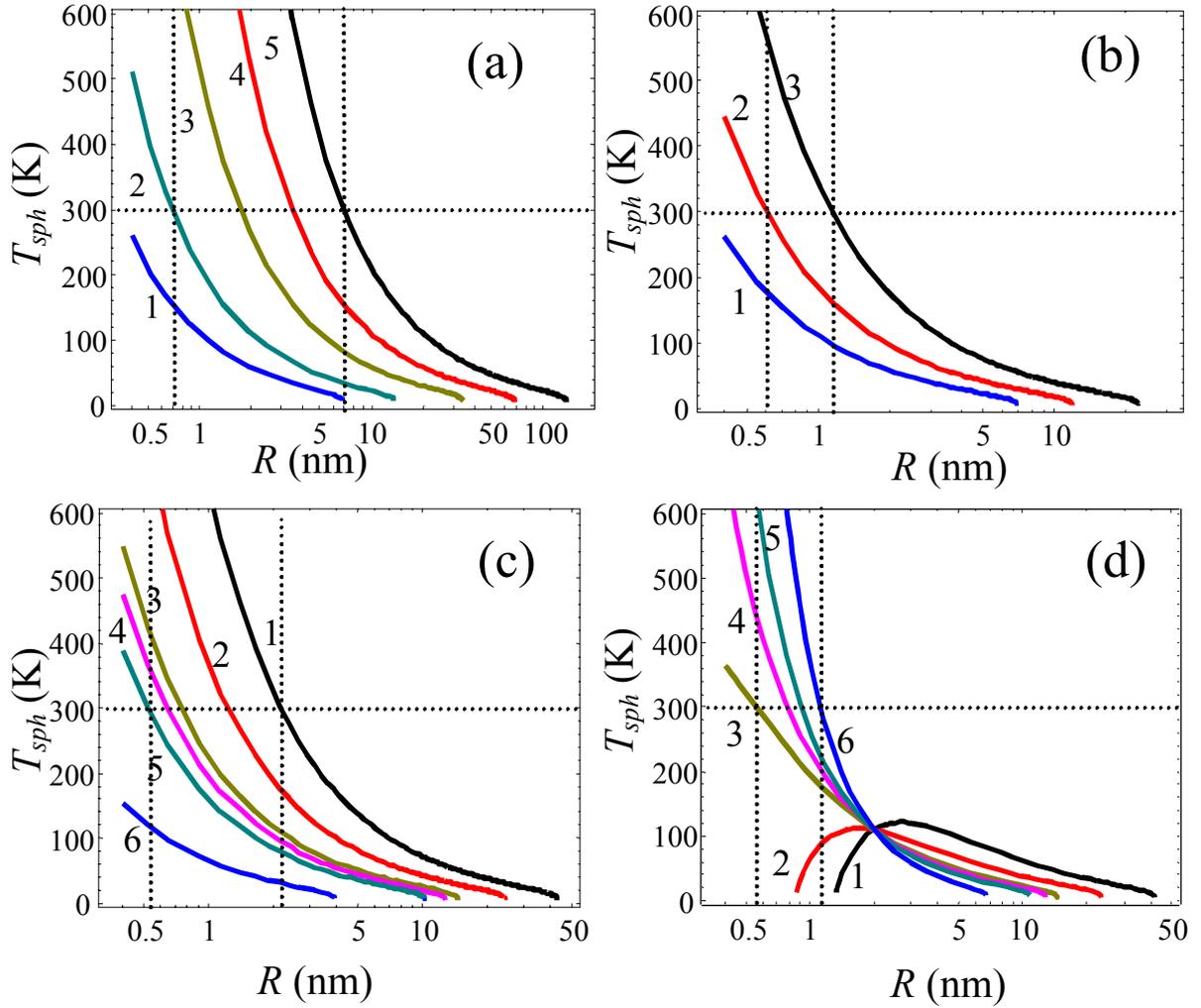

**FIG. 11.** Ferroelectric phase transition temperature $T_{sph}$ vs. the sphere radius $R$ for KTaO$_3$ material parameters $T_q = 55\,\text{K}$, $T_0 = 13\,\text{K}$, Curie-Weiss constant $C_{CW} = 5.6 \times 10^4\,\text{K}$, $Q_{11} + 2Q_{12} = 0.041\,\text{m}^4/\text{C}^2$; gradient coefficient $g = 10^{-10}\,\text{V m}^3/\text{C}$ in SI units and $n_d = 4\pi/3$. **(a)** $|\lambda_g| \to +\infty$, $R_\lambda = 0$ and negative surface stress values $\mu$: -1; -2; -5; -10; -20 N/m (curves 1, 2, 3, 4, 5); **(b)** $\mu = 1\,\text{N/m}$, $R_\lambda = 0$ and negative lengths $\lambda_g = -0.8;\,-0.6;\,-0.4\,\text{nm}$ (curves 1, 2, 3); **(c)** $\mu = -2\,\text{N/m}$, $R_\lambda = 0$ and lengths $\lambda_g = -0.4;\,-1;\,-10;\,10;\,3;\,1\,\text{nm}$ (curves 1, 2, 3, 4, 5, 6); **(d)** $\mu = -2\,\text{N/m}$, $R_\lambda = +2\,\text{nm}$ and $\lambda_g = -0.4;\,-1;\,-10;\,10;\,3;\,1\,\text{nm}$ (curves 1, 2, 3, 4, 5, 6).

It is clear from Figs. 11, that the effect of the ferroelectricity appearance in spherical nanoparticles of incipient ferroelectrics is possible for negative intrinsic surface stress coefficient $\mu$ that stretches the particle, since $(Q_{11} + 2Q_{12}) > 0$ (see plots (a, c, d)). At positive $\mu$ ferroelectric phase appears at negative length $\lambda_S(R)$ (see plot (b)), the latter being possible for the cases depicted in

Fig.8. Note, that $\lambda_S(R) < 0$ for arbitrary radiuses at negative length $\lambda_g$ and radius $R_\lambda$, which is achieved under the conditions $a_1^S < 0$ and $\mu(2q_{12}^S + q_{11}^S) < 0$.

The prediction of size-induced ferroelectricity in KTaO$_3$ nanospheres of radius less then 1-5 nm (see vertical lines in Figs.11) at room temperatures (see horizontal lines in Figs.11) and aforementioned special conditions could be interesting. However, it is difficult to observe in comparison with nanowires, where analogous effect is expected at room temperatures and radius 5-20 nm. The difference is related with the absence of depolarization field in nanowires in contrast to nanospheres with depolarization factor $n_d$ chosen equal to $4\pi/3$.

Note, that Abel et al have found that the hydrostatic pressure $p$ higher than $2 \times 10^9$ Pa does not induces any ferroelectric phase in bulk KTaO$_3$. Moreover, reciprocal susceptibility increases, proving paraelectric phase stability [45]. This result is clear, since electrostriction coefficients combination $(Q_{11} + 2Q_{12})$ is positive and polarization gradient influence can be neglected in the bulk sample, so the positive value $(Q_{11} + 2Q_{12})p$ only suppress ordered state appearance. This experimental fact confirms our conclusion about the absence of ferroelectricity at compressive surface stress, when $(Q_{11} + 2Q_{12})\mu/R > 0$.

Allowing for the facts that proposed theoretical approach describes available experimental data [6, 10, 11] rather well (see Figs.4 and 10), we would like to underline that simultaneous consideration of intrinsic surface stress, depolarization effects and polarization gradient are the key for the adequate description of size-induced phase transitions in ferroelectric nanoparticles.

The polarization gradient influence manifests itself via the transition temperature dependence on $R_S$, where $R_S$ is the bulk material correlation radius at zero temperature. It is worth to note, that in the majority of figures (see Figs.3, 5, 9) we represented the transition temperature via the ratio $R/R_S$. For the most of cases essential increase of transition temperature corresponds to the radiuses less that several $R_S$ even at large $R_\mu$ value, related with intrinsic surface stress.

## 6. Discussion

The properties of magnetic and elastic nanoparticles could be considered in the same way as it has been shown in details for ferroelectric nanoparticles. Under the favourable conditions the approximate formulae (4) can be applied for calculation of transition temperature and phase diagrams of all primary ferroics as it was declared in section 2.

Really, in the case of infinite cylindrical magnetic nanoparticle with order parameter aligned along its axis, depolarization factor is zero and thus inner field $E_3^d$ is absent. Gradient contribution into the renormalization of $a_{ij}(T)$ can be estimated as $g/(\lambda_S R) \cong 10 - 0.1$ CGSM units for mono-domain ferromagnetic nanoparticles with radius of curvature $R = 5-50$ nm and gradient coefficient square root $\sqrt{g} \sim 50$ nm and characteristic length $\lambda_S = 50-500$ nm [22], [36]. The estimation of the surface stress contribution to the renormalization of $a_{ij}(T)$ gives $2Q_{lkij}L_{lk}\frac{\mu}{R} \cong 10^2 - 10$ CGSM units at $R = 5-50$ nm, $L_{lk} \sim 1$ and $\mu \cong 5 \cdot 10^4$ din/cm for bulk magnetostriction coefficients $Q_{lkij} \sim 10^{-9}\, cm^3/erg$ typical for rare-earth alloys [46]. Thus, the striction renormalization may be comparable or essentially larger than the aforementioned gradient contribution for radiuses 5nm<R<50nm, making Eqs.(3-4) valid for magnetic nanorods. The estimations had shown that in small nanoparticles magnetization could appear when it is absent in the bulk, explaining qualitatively experimental results [5].

The estimations made in the end of section 2 have shown the possibility of ferroelectricity appearance in the incipient ferroelectrics nanoparticles that has been confirmed by rigorous calculations in sections 4 and 5. The preferable conditions for this new phenomenon observation could be the following:

(i) The best nanoparticle shape is long nanorod with radius less than several tens of nm that provides depolarization field vanishing and strong surface stress effect. For incipient ferroelectric nanorods of perovskite symmetry (KTaO$_3$ or SrTiO$_3$) effect is possible even at room temperature if $\mu Q_{12} < 0$, i.e. when $\mu > 0$ since $Q_{12} < 0$. Additional desirable condition is the negative length $\lambda_S$, appeared even at positive extrapolation length $\lambda_g > 0$ when nanoparticle radius $R < R_\lambda$, the latter being possible if $\mu q_{12}^S < 0$.

(ii) For spherical nanoparticles the effect of the ferroelectricity appearance is possible at radiuses less than several nm and it is difficult to observe at room temperatures in contrast to nanorods. The difference is related with the absence of depolarization field in nanowires in contrast to nanospheres. Hypothetically size-driven ferroelectric phase transition in nanospheres is possible if $\mu(Q_{11} + 2Q_{12}) < 0$, i.e. when $\mu < 0$, since $(Q_{11} + 2Q_{12}) > 0$ for perovskites. Another possibility is the change of the length $\lambda_S$ sign at some value of the nanoparticle radius that is achieved when $\mu(2q_{12}^S + q_{11}^S) < 0$.

The experimental justification of the theoretical forecast is extremely desirable.

**Appendix A. Rochelle Salt nanorod switching**

The RS nanorod switching between two plain electrodes at applied homogeneous electric field $E_0$ can be calculated within the framework of Landauer model [47] for nanodomain nucleation. The excess of electrostatic energy appeared after the semi-ellipsoidal domain nucleation is

$$\Phi_n(r,l) \cong \pi\psi_S\, lr\left(\frac{r}{l} + \frac{\arcsin\sqrt{1-r^2/l^2}}{\sqrt{1-r^2/l^2}}\right) - \frac{4}{3}\pi r^2 l\, P_{VS} E_0 + \frac{4\pi P_{VS}^2}{3\varepsilon_0 \varepsilon_{11}} n_D(r,l) r^2 l, \quad (A.1)$$

where $r$ is domain radius, $l$ is its length (see Fig.1A(a)).

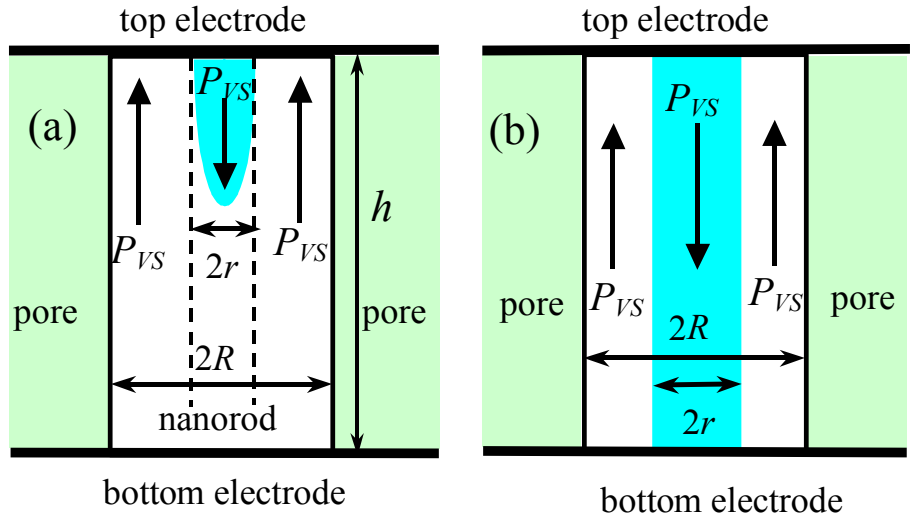

FIG.1A. Semi-ellipsoidal domain nucleation (a) and cylindrical domain, grown through the nanorod (b).

The first term in Eq.(A.1) is the domain wall energy, at that the energy density $\psi_S = 4P_{VS}^3\sqrt{2ga_{11}}/3$ [36] should be calculated from the renormalized free energy expansion (18) at $P_{SV}(R, E_0 = 0) \approx \sqrt{\alpha_T(T_{cyl}(R)-T)/a_{11}}$. Note, that the domain wall could be approximated infinitely thin allowing for the small values of gradient coefficient $g$ and experimentally measured domain wall halfwidth $w \le 0.6$ nm [36]. The second term is the interaction with external field; the third one is Landauer depolarization field energy [47] that depends on dielectric anisotropy factor $\gamma = \sqrt{\varepsilon_{33}/\varepsilon_{11}}$ and effective dielectric permittivity $\kappa = \sqrt{\varepsilon_{33}\varepsilon_{11}}$. The exact expression for depolarization factor $n_D(r,l)$ is well known: $n_D(r,l) = \dfrac{(r\gamma/l)^2}{(r\gamma/l)^2 - 1}\left(1 - \dfrac{\mathrm{arctg}\left(\sqrt{(r\gamma/l)^2 - 1}\right)}{\sqrt{(r\gamma/l)^2 - 1}}\right)$.

Minimizing the free energy (A.1) on domain sizes, one obtains the nucleus radius $r_n$, length $l_n$ and activation energy $E_a = \Phi_n(r_n, l_n)$. Under the condition $\gamma r/l < 1$, one derives that

$r_n \approx 5\pi\psi_S/(16 P_{VS} E_0)$ and $E_a \approx \pi\psi_S^2 l_n/(P_{VS} E_0)$, so $r_n$ and $E_a$ are inversely proportional to the applied electric field $E_0$.

When the length $l_n$ exceeds the nanorod length $h$ (see Fig.1A(b)) one should use the excess of electrostatic energy appeared after the cylindrical nanodomain *vertical* growth through the particle:

$$\Phi_n(r) = 2\pi r h \psi_S - \pi r^2 h P_{VS} E_0 \qquad (A.2)$$

Minimizing the free energy (A.2) on domain radius, one obtains that the nucleus radius $r_n = \psi_S/(P_{VS} E_0)$ and activation energy $E_a = F_n(r_n) = \pi\psi_S^2 h/(P_{VS} E_0)$ are inversely proportional to the applied bias $E_0$. For the case temperature dependent activation electric field $E_{cr}^a$ can be explicitly found from the condition $E_a = k_B T$, namely $E_{cr}^a = \pi\psi_S^2 h/(P_{VS} k_B T)$.

For the general case of arbitrary (in particular not very prolate) semi-ellipsoidal nucleus the activation energy depends on applied bias more complexly than $\sim E_0^{-1}$, thus the thermal activation field $E_{cr}^a$ should be estimated numerically from the equation $E_a(E_{cr}^a) = k_B T$. Usually the activation field $E_{cr}^a$ is closely related with the coercive field $E_C$ [20].

The nucleation time $\tau_n$ can be calculated in accordance with the activation law [48]:

$$\tau_n = \tau_0 \exp(E_a/k_B T), \qquad (A.3)$$

where characteristic times $\tau_0 \sim 10^{-12}$ s. At applied electric fields $E_0 \gg E_{cr}^a$ nucleation time $\tau_n$ is rather small in comparison with the nanorod switching time $\tau_S$ that is mainly determined as the time of *domain wall motion* and so is described by the power law [20]:

$$\tau_S = C_m/E_0^n. \qquad (A.4)$$

Here $n$ is the fitting parameter that usually belongs to the range $0 < n \leq 1$, fitting constant $C_m$ is related with the domain wall mobility [49].

Using parameters $g \approx 9 \cdot 10^{-11}$ m$^3$/F [36], $\alpha_T \approx 7.74 \cdot 10^7$ m/F·K and $a_{11} = 3.36 \cdot 10^{15}$ SI units (estimated from bulk polarization of RS crystal at $T_0 = 0\,°$C is $P_S = \sqrt{\alpha_T T_0/a_{11}} = 0.25\,\mu$C/cm$^2$), we calculated that $P_{VS}(R, E_0 = 0) \approx 0.28\,\mu$C/cm$^2$ at $T$=21°C and $T_{cyl} = 80\,°$C (see Fig.4), thus domain wall energy $\psi_S(R) = 2.33 \cdot 10^{-5}$ J/m$^2$. Also we estimated that $l_n \ll h$ at $E_0 \geq 2$ kV/cm and $h = 500$ nm. So, nuclei have rather semi-ellipsoidal shape than cylindrical one.

Measured in Ref.[6] and calculated from Eqs.(A.1)-(A.4) nanorod switching time $\tau_S$, nucleation time $\tau_n$, corresponding nucleus radius $r_n$, length $l_n$ and activation energy $E_a$ via applied electric field are shown in Figs.2A.

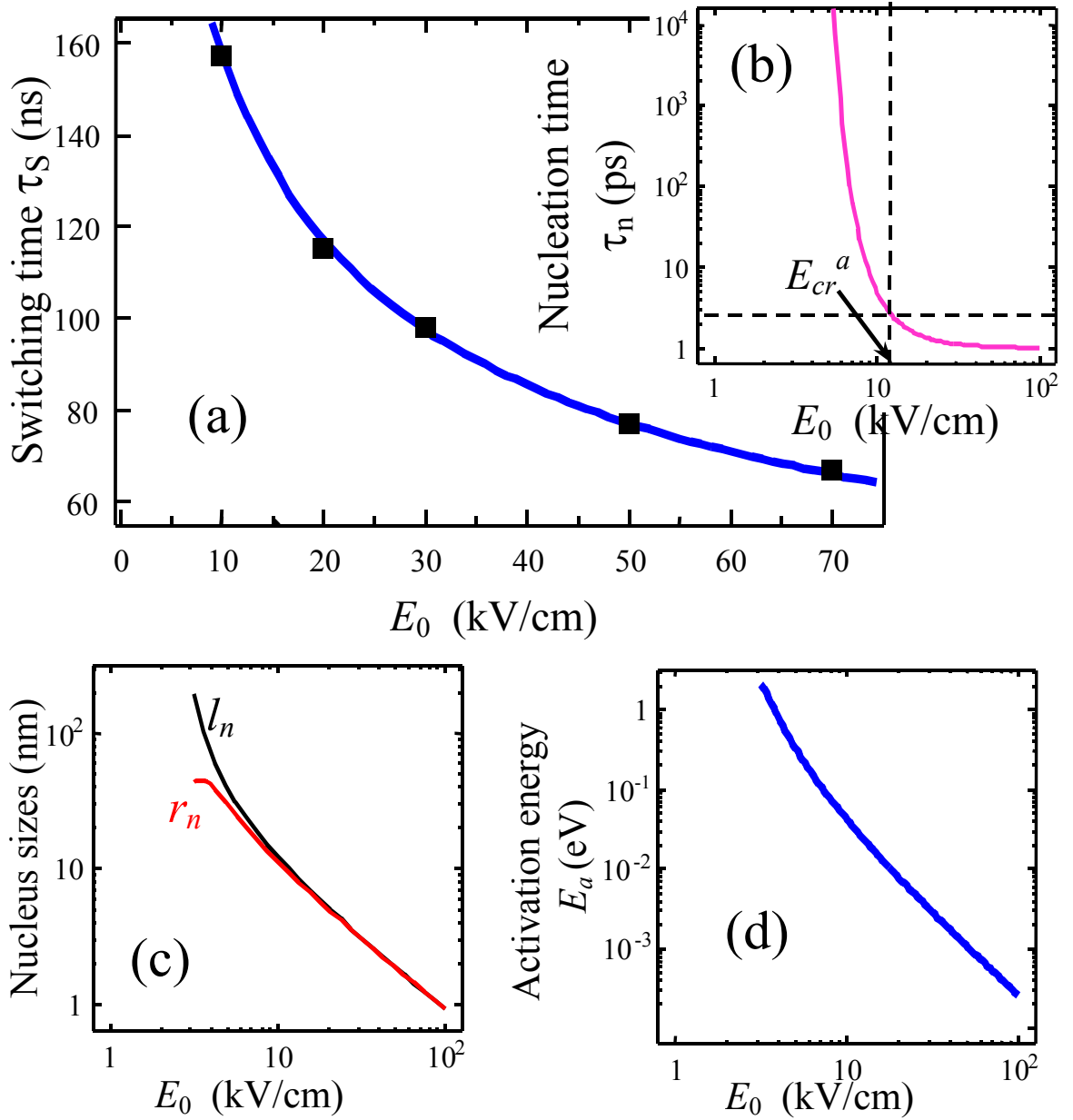

FIG.2A. (a) Switching time $\tau_S$ via applied electric field $E_0$. Squares are experimental data from Ref.[6] measured at 21°C; solid curves are our fitting (A.4) at $C_m \approx 570$ ns·cm/kV and $n = 0.5$. Corresponding nucleation time $\tau_n$ (b), nucleus sizes $r_n$, $l_n$ (c) and activation energy $E_a$ (d) field dependences are calculated from Eqs.(A.1) and (A.3) at $\tau_0 \approx 1$ ps, $\psi_S = 2.33 \cdot 10^{-5}$ J/m² and $P_{VS} \approx 0.28$ μC/cm².

It is clear from Fig.2A(b) that the activation field $E_{cr}^a \approx 13$ kV/cm, the value is in a reasonable agreement with experimentally measured coercive field value $E_C \approx 13.6$ kV/cm [6]. This could denote the fact of thermal intrinsic switching in the nanorod (thermal fluctuations $\sim k_B T$ cause rapid numerous nanodomain nucleation) in contrast to the switching of bulk sample at coercive fields about 0.2 kV/cm that believed to be extrinsic allowing for growth imperfections.

Small values of nucleus sizes $r_n \approx l_n$ less than 5 nm appeared at applied field $E_0 > 10\,\text{kV/cm}$ (see plot (c)) in comparison with nanorod ones ($R = 15\,\text{nm}$ and $h = 500\,\text{nm}$) explain the difference between the relatively small calculated values of $\tau_n$ and high $\tau_S$: rapid domain nucleation is followed by the slow domain wall lateral and vertical growth. Since $h \gg R$ the wall growth is mainly vertical, and so its 1D motion with velocity $\vartheta$ can be described from the kinetic energy conservation law $m_w \vartheta^2 / 2 - q_w E_0 h \cong const$. Obtained expression for velocity $\vartheta \sim \sqrt{2 q_w E_0 h / m_w}$ and switching time $\tau_S \cong h/\vartheta \sim \sqrt{h m_w / 2 q_w E_0}$ clarify the power $n = 0.5$ in Eq.(A.4) obtained from our fitting.

**Appendix B. Transition temperature in nanospheres**

The transcendental equation for the transition temperature $T_{sph}(R)$ acquires the form:

$$a_R + \frac{\dfrac{3 g n_d}{R^2 (n_d + a_R)} \left( R \sqrt{\dfrac{n_d + a_R}{g}} \cosh\left( R \sqrt{\dfrac{n_d + a_R}{g}} \right) - \sinh\left( R \sqrt{\dfrac{n_d + a_R}{g}} \right) \right)}{\lambda_S \sqrt{\dfrac{n_d + a_R}{g}} \cosh\left( R \sqrt{\dfrac{n_d + a_R}{g}} \right) + \left(1 - \dfrac{\lambda_S}{R}\right) \sinh\left( R \sqrt{\dfrac{n_d + a_R}{g}} \right)} = 0 \quad (\text{B.1})$$

Here $a_R(T,R) = a_1(T) + (Q_{11} + 2Q_{12})\dfrac{4\mu}{R}$ and $a_1(T) = \alpha_T(T - T_C^b)$ for conventional ferroelectrics. Under the condition $\varepsilon_e < 10$, typical for nanopowder in air or soft gel/liquid matrix the approximation $n_d + a_R \approx n_d$ is valid with high accuracy. Since typically $\exp(-R/R_d) \ll 1$ for nanospheres of radius $R > 0.5 - 5\,\text{nm}$ we derived from Eq.(B.1) that $T_{sph}(R) \approx T_C^b \left( 1 - \dfrac{R_\mu}{R} - \dfrac{3 R_S^2}{\lambda_g + R_d} \cdot \dfrac{1 - R_\lambda / R}{R - R_d R_\lambda / (\lambda_g + R_d)} \right)$.

Under the absence of depolarization field (e.g. when $\varepsilon_e \gg 100$ and/or free surface charges provide almost perfect screening) transcendental equation for the transition temperature $T_{sph}(R)$ acquires the form [16]:

$$\lambda_S(R) \sqrt{-\dfrac{a_R(T,R)}{g}} \cos\left( R \sqrt{-\dfrac{a_R(T,R)}{g}} \right) + \left(1 - \dfrac{\lambda_S(R)}{R}\right) \sin\left( R \sqrt{-\dfrac{a_R(T,R)}{g}} \right) = 0 \quad (\text{B.2})$$

Here $a_R(T,R) = a_1(T) + (Q_{11} + 2Q_{12})\dfrac{4\mu}{R}$ and $a_1(T) = \alpha_T(T - T_C^b)$ for conventional ferroelectrics. Pade approximations of Eq.(B.2) solution could be rewritten as:

$$T_{sph}(R) \approx \begin{cases} T_C^b - (Q_{11} + 2Q_{12})\dfrac{4\mu}{\alpha_T R} - \dfrac{g}{\alpha_T} \dfrac{3}{R \lambda_S(R) + 3R^2/\pi^2}, & \lambda_S(R) \geq 0, \\[2ex] T_C^b - (Q_{11} + 2Q_{12})\dfrac{4\mu}{\alpha_T R} - \dfrac{g}{\alpha_T} \left( \dfrac{3}{R \lambda_S(R)} - \dfrac{1}{\lambda_S^2(R)} \right), & \lambda_S(R) < 0. \end{cases} \quad (\text{B.3})$$

**References**

1. V.K. Wadhawan. *Introduction to ferroic materials*. Gordon and Breach Science Publishers (2000).

2. E. Roduner. *Nanoscopic Materials. Size-dependent phenomena*. RSC Publishing (2006).

3 Michael G. Cottan. *Linear and nonlinear spin waves in magnetic films and super-lattices*. World Scientific, Singapore (1994).

4. D. R. Tilley, *Finite size effects on phase transitions in ferroelectrics*. in: *Ferroelectric Thin Films*, ed. C. Paz de Araujo, J. F.Scott, and G. W. Teylor (Gordon and Breach, Amsterdam, 1996) 11.

5. Y. Nakal, Y. Seino, T. Teranishi, M. Miyake, S. Yamada, H. Hori, Physica B, **284**, 1758 (2000).

6. D. Yadlovker and S. Berger, Uniform orientation and size of ferroelectric domains. *Phys. Rev.* B **71**, 184112 (2005).

7. A.N. Morozovska, E.A. Eliseev, and M.D. Glinchuk, Ferroelectricity enhancement in confined nanorods: Direct variational method, Phys. Rev. B **73**, 214106 (2006).

8. A.N. Morozovska, E.A. Eliseev, and M.D. Glinchuk, Physica B. **387**, 358 (2007).

9. M. H. Frey and D. A. Payne, Grain-size effect on structure and phase transformations for barium titanate *Phys. Rev.* B 54, 3158- 3168 (1996).

10. Z. Zhao, V. Buscaglia, M. Viviani, M.T. Buscaglia, L. Mitoseriu, A. Testino, M. Nygren, M. Johnsson, and P. Nanni, Grain-size effects on the ferroelectric behavior of dense nanocrystalline BaTiO3 ceramics. *Phys. Rev.* B **70**, 024107-1-8 (2004).

11. E. Erdem, H.-Ch. Semmelhack, R. Bottcher, H. Rumpf, J. Banys, A.Matthes, H.-J. Glasel, D. Hirsch and E. Hartmann. Study of the tetragonal-to-cubic phase transition in PbTiO3 nanopowders. J. Phys.: Condens. Matter **18** 3861–3874 (2006).

12. F. Wiekhorst, E. Shevchenko, H. Weller, and J. Kotzler, Anisotropic superparamagnetism of monodispersive cobalt-platinum nanocrystals. *Phys. Rev.* B **67**, 224416 (2003)

13. M. Respaud, J. M. Broto, H. Rakoto, and A. R. Fert. Surface effects on the magnetic properties of ultrafine cobalt particles. *Phys. Rev.* B, **57**, 2925-2935 (1998)

14. N. A. Pertsev, A. K. Tagantsev, and N. Setter. Phase transitions and strain-induced ferroelectricity in SrTiO$_3$ epitaxial thin films. Phys. Rev. B **61**, №2, R825 - 829 (2000).

15. W.L. Zhong, Y. G. Wang, P.L. Zhang, and D. B. Qu. Phenomenological study of the size effect on phase transition in ferroelectric particles. *Phys. Rev.* B **50**, 698 (1994).

16. C. L. Wang and S. R. P. Smith, Landau theory of the size-driven phase transition in ferroelectrics. *J. Phys.: Condens. Matter* **7**, 7163 (1995).

17. I. Rychetsky and O. Hudak, The ferroelectric phase transition in small spherical particles. *J. Phys.: Condens. Matter* **9**, 4955 (1997).

18. J. Zhang, Zh. Yin, M.-Sh. Zhang, and J. F. Scott, Size-driven phase transition in stress-induced ferroelectric thin films. *Solid State Communications* **118**, 241 (2001).

19. L.D. Landau and E.M. Lifshits, *Electrodynamics of Continuous Media*, (Butterworth Heinemann, Oxford, 1980).

20. M. E. Lines and A. M. Glass, *Principles and Applications of Ferroelectrics and Related Phenomena* (Clarendon Press, Oxford, 1977).

21. Ch. Kittel., *Introduction to solid state physics*, London, Chapmen & Hall (1956).

22. M. I. Kaganov and A. N. Omelyanchouk, Contribution to the phenomenological theory of a phase transition in a thin ferromagnetic plate. Zh. Eksp. Teor. Fiz. **61**, 1679 (1971) [Sov. Phys. JETP **34**, 895 (1972)].

23. R. Kretschmer and K. Binder, Surface effects on phase transition in ferroelectrics and dipolar magnets. *Phys. Rev*. B **20**, 1065 (1979).

24. L.D. Landau and E.M. Lifshitz, Theory of Elasticity. Theoretical Physics, Vol. 7 (Butterworth-Heinemann, Oxford, U.K., 1998).

25. O. Song, C. A. Ballentine, and R. C. O'Handley. Giant surface magnetostriction in polycrystalline Ni and NiFe films. *Appl. Phys. Lett.* **64** (19), 2593 (1994).

26. M.D. Glinchuk, and A.N. Morozovska, The internal electric field originating from the mismatch effect and its influence on ferroelectric thin film properties. J. Phys.: Condens. Matter **16**, 3517 (2004).

27. A.M. Bratkovsky, and A.P. Levanyuk. Smearing of Phase Transition due to a Surface Effect or a Bulk Inhomogeneity in Ferroelectric Nanostructures**.** *Phys. Rev. Lett*. **94**, 107601 (2005).

28. V.I. Marchenko, and A.Ya. Parshin, About elastic properties of the surface of crystals. Zh. Eksp. Teor. Fiz. **79** (1), 257-260 (1980), [Sov. Phys. JETP **52**, 129-132 (1980)].

29. V.A. Shchukin, D. Bimberg, Spontaneous ordering of nanostructures on crystal surfaces. Rev. Mod. Phys. **71**(4), 1125-1171 (1999).

30. K. Uchino, E. Sadanaga, and T. Hirose, Dependence of the crystal structure on particle size in barium titanate. *J. Am. Ceram. Soc.* **72**, (8) 1555-1558 (1989).

31. W. Ma, M. Zhang, and Z. Lu, A study of size effects in $PbTiO_3$ nanocrystals by Raman spectroscopy. Phys. Stat. Sol. (a) **166**, 811-815 (1998).

32. Z. H. Zhou, X. S. Gao, and John Wang, K. Fujihara and S. Ramakrishna, V. Nagarajan, Giant strain in $PbZr_{0.2}Ti_{0.8}O_3$ nanowires, *Appl. Phys. Lett* **90**, 052902-1-3 (2007).

33. N. A. Pertsev, A. G. Zembilgotov, and A. K. Tagantsev, Effect of Mechanical Boundary Conditions on Phase Diagrams of Epitaxial Ferroelectric Thin Films. *Phys. Rev. Lett*. **80**, 1988 (1998).

34. M.D.Glinchuk, E.A.Eliseev, and V.A.Stephanovich, The depolarization field effect on the thin ferroelectric films properties. Physica B, **332**, 356 (2002).

35. M. D.Glinchuk, A. N. Morozovska, E. A. Eliseev, Ferroelectric thin films phase diagrams with self-polarized phase and electret state. *J. Appl. Phys*. **99**(11), 114102 (2006).

36. G Catalan, J F Scott, A Schilling and J M Gregg. Wall thickness dependence of the scaling law for ferroic stripe domains. *J. Phys.: Condens. Matter* **19**, 022201 (2007).

37. E.D. Mishina, N.E. Sherstyuk, V.O. Valdner, A.V. Mishina, K.A. Vorotilov, V.A. Vasiliev, A.S. Sigov, M.P. De Santo, E. Cazzanelli, R. Barberi, and Th. Rasing, Nonlinear optical and microRaman diagnostics of thin films and nanostructures of ABO3 ferroelectrics, Solid State Physics, **48**(6), 1133 (2006).

38. F. D. Morrison Y. Luo, I. Szafraniak, V. Nagarajan, R. B. Wehrspohn, M. Steinhart, J. H. Wendroff, N. D. Zakharov, E. D. Mishina, K. A. Vorotilov, A. S. Sigov, S. Nakabayashi, M. Alexe, R. Ramesh, and J. F. Scott, Ferroelectric nanotubes. *Rev. Adv. Mater. Sci*. **4**, 114 (2003).

39. J. H. Barrett, Dielectric Constant in Perovskite Type Crystals. Phys. Rev. **86**, 118 (1952).

40. H. Uwe and T. Sakudo, Raman-scattering study of stress-induced ferroelectricity in $KTaO_3$. Phys. Rev. B **15**, №1, 337 - 345 (1977).

41. M. D. Glinchuk and A. N. Morozovska, Effect of Surface Tension and Depolarization Field on Ferroelectric Nanomaterials Properties. *Phys. Stat. Sol.* (b) **238**, 81 (2003).

42. D.A. Scrymgeour and V.Gopalan, A.Itagi, A.Saxena, and P.J. Swart. Phenomenological theory of a single domain wall in uniaxial trigonal ferroelectrics: Lithium niobate and lithium tantalite. *Phys. Rev*. B **71**, 184110 (2005).

43. G. B. Stephenson, and K. R. Elder. Theory for equilibrium 180° stripe domains in $PbTiO_3$ films. *J. Appl. Phys*. **100**, 051601 (2006)

44. K. Ishikava, K. Yoshikawa, N. Okada, Size effect on the ferroelectric phase transition in PbTiO3 ultrafine particles. *Phys. Rev*. B **37**, 5852-5857 (1988).

45. W. R. Abel, Effect of Pressure on the Static Dielectric Constant of $KTaO_3$**,** Phys. Rev. B **4**, № 8, 2696 - 2701 (1971).

46. Ce Wen, Ming Li. Calculations of giant magnetoelectric effects in ferroic composites of rare-earth alloys and ferroelectric polymers. *Phys. Rev*. B **63**, 144415-1-9 (2001).

47. R. Landauer, Electrostatic Considerations in $BaTiO_3$ Domain Formation during Polarization Reversal. *J. Appl. Phys*. **28**, 227 (1957).

48. A.N. Morozovska, S.V. Kalinin, and E.A. Eliseev, Domain nucleation and hysteresis loop shape in piezoresponse force spectroscopy. *Appl. Phys. Lett.* **89**, 192901 (2006).

49. P. Paruch, T. Giamarchi, T. Tybell, and J.-M. Triscone. Nanoscale studies of domain wall motion in epitaxial ferroelectric thin films**.** *J. Appl. Phys*. **100**, 051608 (2006)